\documentclass{article}

\usepackage{arxiv}

\usepackage[utf8]{inputenc} 
\usepackage[T1]{fontenc}    
\usepackage{hyperref}       
\usepackage{url}            
\usepackage{booktabs}       
\usepackage{amsfonts}       
\usepackage{nicefrac}       
\usepackage{microtype}      
\usepackage{graphicx}
\usepackage{doi}
\usepackage[version=3]{mhchem} 
\usepackage{siunitx}
\usepackage{extarrows} 
\usepackage{float}
\usepackage{amsmath}
\usepackage[title]{appendix}
\usepackage[superscript,biblabel]{cite}
\usepackage[font={small}]{caption}

\title{Alkalinity Concentration Swing for Direct Air Capture of Carbon Dioxide}

\author{\\ 
    $^\dagger$John A. Paulson School of Engineering and Applied Sciences\\
	Harvard University\\
	Cambridge, Massachusetts, 02138, USA \\
	\And
	\\
	$^\mathsection$Dept. of Earth and Planetary Sciences\\
	Harvard University\\
	Cambridge, Massachusetts, 02138, USA\\
}

\date{  April 2021 \\
        rinberg@g.harvard.edu \\
        $^*$\emph{These authors contributed equally to this work.}}

\renewcommand{\headeright}{Preprint}

\renewcommand{\undertitle}{\textbf{Anatoly Rinberg,$^{*\dagger}$ Andrew M. Bergman,$^{*\dagger}$ Daniel P. Schrag,$^{\mathsection\dagger}$ Michael J. Aziz$^\dagger$}}
\renewcommand{\shorttitle}{}

\hypersetup{
pdftitle={Alkalinity Concentration Swing for Direct Air Capture of Carbon Dioxide},
pdfsubject={},
pdfauthor={Anatoly ~Rinberg, Andrew M.~Bergman, Daniel P.~Schrag, Michael J.~Aziz},
pdfkeywords={carbon dioxide removal, direct air capture, carbon dioxide, reverse osmosis,  capacitive deionization},
}

\begin{document}
\maketitle

\begin{abstract}
{We describe a new principle — the Alkalinity Concentration Swing (ACS) — for direct air capture of carbon dioxide driven by concentrating an alkaline solution that has been exposed to the atmosphere and loaded with dissolved inorganic carbon. Upon concentration, the partial pressure of carbon dioxide increases, allowing for extraction and compression. We find that higher concentration factors result in proportionally higher outgassing pressure, and higher initial alkalinity concentrations at the same concentration factor outgas a higher concentration of \ce{CO2} relative to the feed solution. We examine two desalination technologies, reverse osmosis and capacitive deionization, as possible implementation for the ACS, and evaluate two simplified corresponding energy models. We compare the ACS to incumbent technologies and make estimates on water, land, and energy requirements for capturing one million tonnes of \ce{CO2} per year. We find that estimates for the lower end of the energy range for both reverse osmosis and capacitive deionization approaches are lower than or roughly equal to incumbent direct air capture approaches. For most conditions, we find an inverse relationship between the required energy and water processing volume per million tonnes of \ce{CO2}. Realizing the ACS requires a simple alkaline aqueous solvent (e.g. potassium alkalinity carrier) and does not require heat as a driving mechanism. More generally, the ACS can be implemented through industrial-scale desalination approaches, meaning current technology could be leveraged for scale-up.}
\end{abstract}

\newpage

\tableofcontents

\newpage
\section{Introduction}\label{sec:intro}
Removal of carbon dioxide from the atmosphere has been proposed as an essential method for responding to anthropogenic climate change.\cite{minx_negative_2018, fuss_negative_2018} Policymakers and scientists agree that in order to minimize future harm to society — which will be most felt by the world’s most vulnerable populations — priority should be devoted to efforts and technologies that reduce emissions from burning fossil fuels and other sources of greenhouse gases.\cite{ipcc_global_2018} But even after deep decarbonization efforts, some hard-to-avoid emissions will remain, either because they are unacceptable to avoid from a social-justice perspective (e.g. food security constraints) or extremely physically difficult to eliminate within the given timeframe, making some degree of carbon dioxide removal (CDR) necessary.\cite{bergman_case_2021,davis_net-zero_2018} A gigatonne-per-year scale of global CDR will likely be required by the end of the century. However, aiming for larger scales, up to 20 billion tonnes of \ce{CO2} per year (Gt\ce{CO2}/year) as some reports suggest,\cite{nasem_negative_2019} has significant associated moral hazards and ethical considerations, as the promise of future CDR could deter decarbonization in the short term.\cite{lenzi_ethics_2018,anderson_trouble_2016}

\subsection{Carbon dioxide removal}\label{sec:intro:cdr}
Carbon dioxide removal spans a wide range of approaches, each with different associated materials, energy, land, resource, and societal consideration. Biological CDR methods — including reforestation\cite{griscom_natural_2017,anderegg_climate-driven_2020} and soil carbon management\cite{smith_soil_2016} — are projected to be able to achieve gigatonne-scale per year removal,\cite{nasem_negative_2019} though they tend to store carbon in impermanent reservoirs, meaning they are more susceptible to reversals. While these approaches confer co-benefits, such as increasing biodiversity and improving local water and soil quality, they also require significant land area to reach gigatonne scale and may compete with other land-use demands, such as agriculture or conservation objectives.\cite{fuss_negative_2018}

Other approaches that store carbon in a more durable form include bioenergy with carbon capture and storage,\cite{fajardy_can_2017} carbon mineralization processes that remove \ce{CO2} directly out of the air,\cite{mcqueen_ambient_2020} or the addition of alkalinity to oceans, which increases dissolved carbon and ultimately drives the production of carbonate sediments.\cite{harvey_mitigating_2008,rau_global_2018,house_electrochemical_2007,renforth_assessing_2017} Mineralization processes could result in the long-term storage of concentrated \ce{CO2} streams in subsurface formations, products such as concrete, as well as mine tailings and alkaline industrial wastes.\cite{nasem_negative_2019,kelemen_overview_2019}

\subsection{Industrial direct air capture approaches}\label{sec:intro:dac}
An alternate strategy for carbon dioxide removal involves direct air capture (DAC), industrial technologies for separating atmospheric \ce{CO2} directly from the air through chemical or physical processes,\cite{sanz-perez_direct_2016} coupled with sequestration (e.g., storage in a geological reservoir or through mineralization).

One class of approaches is based on solid sorbent technologies that typically use amine-based materials to reversibly bind \ce{CO2}. This process can be cycled many times to capture \ce{CO2} out of ambient air and release a concentrated stream through a thermal\cite{sanz-perez_direct_2016} or moisture swing.\cite{wang_moisture_2011} Alternatively, recent work demonstrated a faradaic electro-swing adsorption system, which uses voltage to regenerate \ce{CO2}.\cite{voskian_faradaic_2019}

Another class of approaches relies on a basic aqueous solution to absorb \ce{CO2} from ambient air. One commercialized process is based on an aqueous potassium hydroxide contactor that absorbs \ce{CO2} directly from air and converts the \ce{CO2} to calcium carbonate; releasing the \ce{CO2} requires heating of calcium carbonate to approximately 900°C.\cite{keith_process_2018} A different approach makes use of an electrochemical swing, which changes the pH of the solution and allows for the release of \ce{CO2} without going through the steps of precipitation and heating.\cite{jin_ph_2020}  

In this paper, we describe a new DAC approach that is based on taking a dilute alkaline aqueous solution that has equilibrated with air, and concentrating it. Concentrating the alkalinity and the dissolved carbon increases the partial pressure of \ce{CO2} in the solution and allows for \ce{CO2} outgassing and extraction. We describe the chemical cycle underlying this approach, evaluate the thermodynamics of the process, and examine two commercially available technologies that are traditionally used for desalination, reverse osmosis and capacitive deionization, to drive the cycle. We then compare the potential advantages of this approach relative to other existing DAC methods and analyze its scale-up feasibility.
\\
\section{The Alkalinity Concentration Swing for direct air capture}\label{sec:ACS_theory}
Our new approach for direct air capture is based on the recognition that an alkaline aqueous solution containing an air-equilibrated concentration of dissolved inorganic carbon (DIC) — the sum of carbonate ion (\ce{CO_3^-^2}), bicarbonate ion (\ce{HCO_3^-}), and dissolved aqueous carbon dioxide (\ce{CO2}(aq)) concentrations in solution — will release \ce{CO2} to the air when that solution is concentrated. After outgassing, if the same alkaline aqueous solution is diluted, \ce{CO2} absorbs from the air and the DIC concentration increases. We use this phenomenon as the core component of the “Alkalinity Concentration Swing” (ACS) cycle for capturing atmospheric \ce{CO2}. In this section we describe the basis for this approach and present a set of idealized steps for realizing the ACS.

\subsection{The equilibrium aqueous carbonate system for varied alkalinity}\label{sec:ACS_theory:eq}

\begin{figure}[h]
 \centering
 \includegraphics[width=\textwidth]{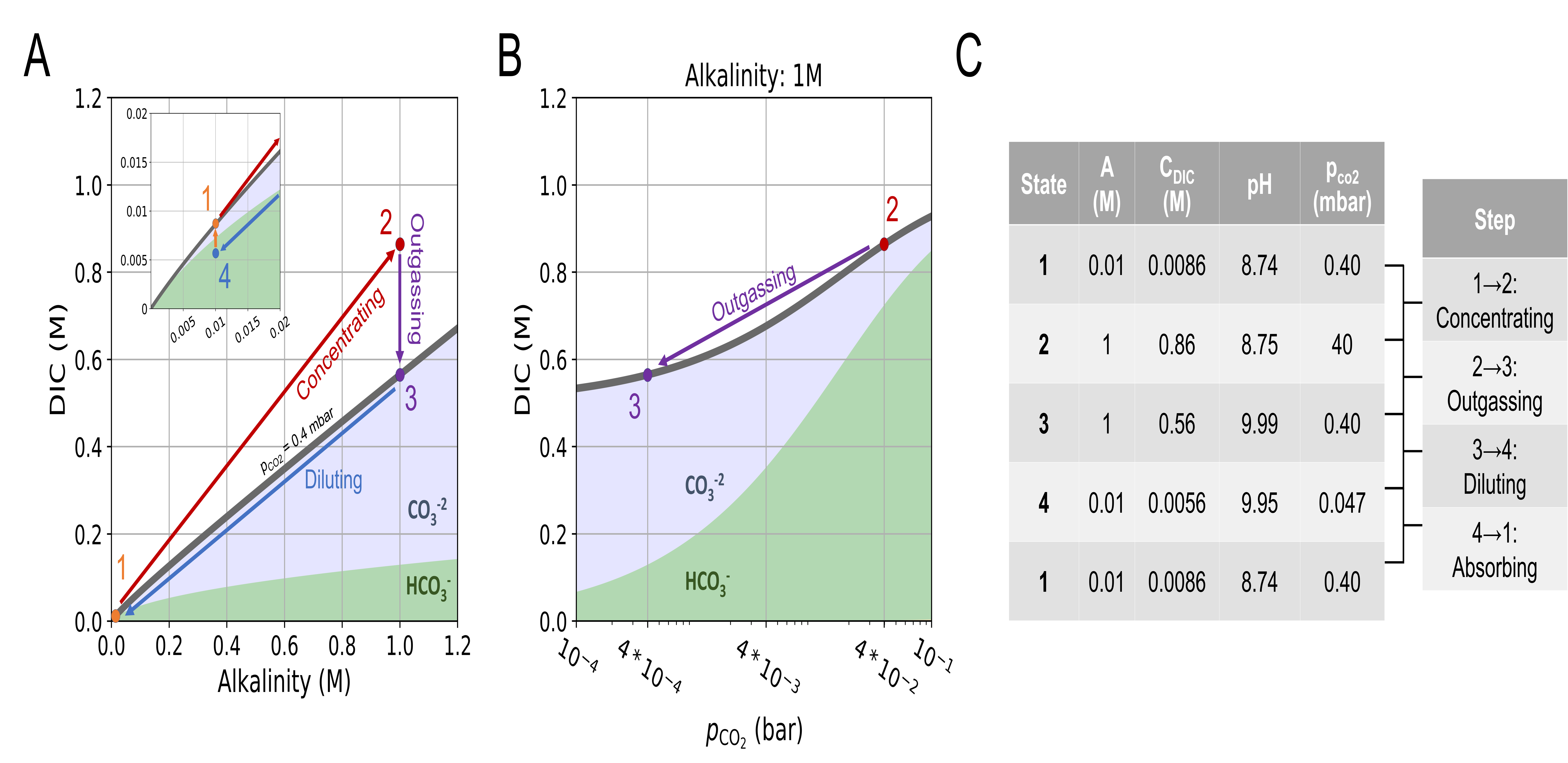}
 \caption{An Alkalinity Concentration Swing Cycle. (A) The gray line plots the dissolved inorganic carbon (DIC) concentration in equilibrium with atmospheric \ce{CO2} ($p_i= 0.4$ mbar) as a function of alkalinity. The red arrow indicates the concentration step of the ACS and plots the trajectory when a solution, initially equilibrated at 0.01 M alkalinity (orange point), is concentrated by a factor of 100 to an alkalinity of 1 M. In the concentrated state (red point), the solution has excess DIC over that in equilibrium with air. The purple arrow indicates the amount of \ce{CO2} that outgasses as the system reaches a new equilibrium at high alkalinity and $p_f = 0.4$ mbar. The remaining blue and orange arrows indicate the dilution and \ce{CO2} absorption steps, which return the system to the initial state. Vertical extent of green and purple lavender regions indicate concentrations of bicarbonate and carbonate, respectively. Inset: The black curve plots the same DIC versus alkalinity relationship as the main plot, but from 0 to 0.02 M alkalinity, showing the transition between a roughly 1:1 alkalinity to DIC relationship at low alkalinity to a 2:1 scaling at higher alkalinity. (B) A plot of DIC as a function of $p_{CO_2}$ at a fixed alkalinity of 1M. The red and purple dots, as well as the purple arrow, correspond to panel A. (C) The alkalinity, $C_{DIC}$, pH, and $p_{CO_2}$ values are listed corresponding to each state in (A).}
 \label{fgr:acs_example}
\end{figure}

The concentration of DIC in equilibrium with atmospheric \ce{CO2} ($p_{CO_2} \approx 0.4$ mbar; equivalent to roughly 400 ppm) depends on the alkalinity — the molar charge difference between the sum of the conservative cations and that of the conservative anions, i.e. ions whose concentrations do not vary with pH. For simplicity, this work assumes alkalinity to be the moles of \ce{K+} ions per liter. As the alkalinity of the solution increases and equilibrium with air is maintained, the pH increases and the amount of DIC increases, but at a decreasing rate. This is due to the transition of the dominant species of DIC from bicarbonate to carbonate as pH increases above $\sim$9.5. Indeed, in very dilute solutions (alkalinity < \SI{1e-2}{M}), the slope of the DIC-alkalinity line is close to unity (Figure \ref{fgr:acs_example}A inset), where each unit of alkalinity, or conservative cations, is balanced by monovalent bicarbonate ions. At alkalinity > \SI{0.1}{M}, the slope is closer to 0.5 (Figure \ref{fgr:acs_example}A main plot) and the charge balance required by this increase in alkalinity is accommodated primarily by the carbonate ion, which is divalent.

The DIC to alkalinity relationship follows from carbonate and aqueous chemistry equilibrium relations, as well as the charge-neutrality condition requiring that the excess charge of conservative cations over conservative anions equal the excess charge of non-conservative anions over non-conservative cations (see Appendix Section A.2). The relative ratios of carbon species in equilibrium is set by the following chemical reactions:
\begin{equation}
    CO_2(gas) \xleftrightarrow{H^{cp}} CO_2(aq)
\end{equation}
\begin{equation}
   CO_2(aq) + H_2O \xleftrightarrow{K_1} H^+ + HCO_{3}^{-} \xleftrightarrow{K_2} 2H^+ + CO_3^{-2}
\end{equation}

The following system of equations determines the relationship between \ce{CO2} partial pressure, alkalinity, and DIC:
\begin{equation}
    [CO_2]_{aq} = H^{cp} p_{CO_2}
\end{equation}
\begin{equation}
    K_1 = [HCO_3^-] [H^+]/[CO_2]_{aq}
\end{equation}
\begin{equation}
    K_2 = [CO_3^{-2}] [H^+]/[HCO_3^-]
\end{equation}
\begin{equation}
    K_w = [H^+][OH^-]
\end{equation}
\begin{equation}
    A  = [HCO_3^-] + 2 [CO_3^{-2}] + [OH^-] - [H^+]
\end{equation}

Here, $A$ is the alkalinity concentration in units of moles per liter. DIC concentration is defined as $C_{DIC}\equiv [CO_2]_{aq} + [HCO_{3}^{-}] +[CO_3^{-2}]$. At a fixed temperature, $K_1$ and $K_2$ vary slightly with the ionic strength\cite{roy_dissociation_1993} and hydrostatic pressure\cite{ludwig_significance_2005} of the solution. Across the studied conditions, these effects modestly enhance the efficiency of the ACS, but are neglected in this study for simplicity; equilibrium constants are fixed for pure water conditions at 20°C and 101.325 kPa: $K_1=$ \SI{9.6e-7}{M}, $K_2=$ \SI{3.4e-10}{M}.\cite{roy_dissociation_1993} Henry's constant, $H^{cp}$, is \SI{0.034}{M/bar} and the water disassociation constant, $K_w$, is \SI{1e-14}{M^2}.
 
The decrease of the ratio of DIC to alkalinity with increasing alkalinity is the principle underlying the ACS. Consider a dilute solution of any strong base initially in equilibrium with air. If that solution is isolated from air and concentrated, for example through the removal of water, both the DIC and the alkalinity increase linearly in proportion to their relative concentrations in the solution. The DIC of a solution with the same corresponding alkalinity, but maintained in equilibrium with air, increases more slowly than that of the concentrated solution. Thus, the concentrated solution is supersaturated and spontaneously outgasses, which allows for the extraction of \ce{CO2}.

This process is depicted in Figure \ref{fgr:acs_example}, in which a solution with an initial alkalinity of 10 mM is concentrated by a factor of 100. When the solution reaches 1 M, $p_{CO_2}$ becomes 40 mbar, which is a hundredfold increase over that in air. When the \ce{CO2} outgasses, restoring equilibrium with air, 0.3 moles of DIC per liter of concentrated solution have been captured (Figure \ref{fgr:acs_example}B). Relative to the initial feed solution, 3 mM of \ce{CO2} or 35\% of the DIC has been outgassed.

\subsection{The Alkalinity Concentration Swing cycle}\label{sec:ACS_theory:cycle}
The following is an idealized description of the ACS based entirely on equilibrium aqueous carbonate assumptions described in Section \ref{sec:ACS_theory:eq} above. Specific methods for implementing the ACS and associated energetics are discussed in Sections \ref{sec:implementing_ACS} and \ref{sec:energy}.

\subsubsection{Step 1\textrightarrow2: Concentrating alkalinity}\label{sec:ACS_theory:cycle:step12}
A solution with initial alkalinity $A_i$ is at equilibrium with the atmosphere at a given partial pressure of \ce{CO2}, $p_i \approx 0.4$ mbar (State 1). The fraction of carbon species in solution and DIC concentration is set by $A_i$ and $p_i$ based on the aqueous carbon chemistry relations described in Section \ref{sec:ACS_theory:eq} above: $C_{DIC,1} = C_{DIC}(A_i,p_i)$ (see Appendix Section A.2 for full derivation).

The system is then closed off from exchange with the atmosphere and the solution is concentrated such that the new effective alkalinity and DIC concentrations are increased by a concentration factor, $\chi$ (Figure \ref{fgr:acs_example}; Concentrating step). Such a concentrating step does not change the absolute number of alkaline carrier ions or DIC molecules in solution, but increases the concentration of both by confining the solutes in a smaller volume. This is equivalent to removing solvent water molecules from solution. The alkalinity and DIC concentrations in the concentrated state are given by: $A_f = A_i \chi$ and $C_{DIC,2} = C_{DIC,1} \chi$, respectively (State 2).

\subsubsection{Step 2\textrightarrow3: \ce{CO2} outgassing}\label{sec:ACS_theory:cycle:step23}
Once the system is in the concentrated state at the higher concentration of alkalinity, $A_f$, the aqueous \ce{CO2} activity increases such that its equilibrium partial pressure rises to $p_{2}$ (State 2). In engineered systems, \ce{CO2} will generally be collected from the concentrated solution by exposing it to a fixed outgassing pressure, $p_f$ that is lower than $p_{2}$ (which we also refer to as $p_{f,max}$).

Exposing the system to $p_f$ in the concentrated state drives the following disproportionation reaction: 

\begin{equation}
    2HCO_{3}^{-} \rightarrow CO_2(aq) + CO_3^{-2} + H_2O
\label{eq:disproportionation}
\end{equation}

 Outgassing occurs as shown in Figure \ref{fgr:acs_example}A-B. The concentration of DIC outgassed as \ce{CO2} with respect to the feed solution as a result of the ACS is given by the following relationship (see Appendix Section B.1 for full derivation):

\begin{equation}
    C_{out}  = C_{DIC}(A_{i},p_i) - \dfrac{A_i}{A_f} C_{DIC}(A_{f},p_f)
\label{eq:outgas}
\end{equation}

The fraction of DIC species outgassed is given by:
\begin{equation}
    f_{out} = C_{out}/C_{DIC}(A_{i},p_i)
\label{eq:fraction}
\end{equation}

The upper limit of $f_{out}$ is 0.5; it occurs only if the initial DIC is entirely made up of bicarbonate ions, and so the alkalinity to DIC ratio is exactly 1:1. If such a system is concentrated to a point where the DIC equilibrium shifts essentially entirely to carbonate at high alkalinity, in the 2:1 alkalinity to DIC regime, then half of the bicarbonate ions are converted to carbonate ions, and the other half become carbon dioxide molecules, which may be collected. In practice, the alkalinity to DIC ratio will fall between 1 and 2.

The maximum pressure at which outgassed \ce{CO2} can be removed from the system is also of interest. Over the range of initial alkalinity between 10-4 and 10 M, the outgassing pressure limit is independent of initial alkalinity and is a direct relationship between the concentrating factor, $\chi$, and the initial pressure, $p_i$, given by (see Appendix Sections A.3 and B.2 for full derivation): 

\begin{equation}
    p_{f,max} \approx p_i \chi
\label{eq:p_max}
\end{equation}

Step 2\textrightarrow3 is concluded once the system has reached its new equilibrium point at $A_f$ and $p_f$, setting a DIC concentration of $C_{DIC,3}$ (State 3).

\subsubsection{Step 3\textrightarrow4: Diluting alkalinity}\label{sec:ACS_theory:cycle:step34}
The next step of the ACS involves returning the concentrated alkalinity, $A_f$, to its initial value. This can be done by recombining the concentrated solution with the removed water from Step 1\textrightarrow2. Alkalinity is diluted by a factor of $1/\chi$ to $A_i$ and DIC is diluted by the same factor giving $C_{DIC,4} = C_{DIC,3}/ \chi$ (State 4).

\subsubsection{Step 4\textrightarrow1: Absorption of atmospheric \ce{CO2}}\label{sec:ACS_theory:cycle:step41}
The final step, which returns the system to State 1, exposes the solution to the atmosphere. Absorption occurs because the dilution step has created a condition with less DIC relative to the concentration in equilibrium with the atmosphere. \ce{CO2} is consumed via the comproportionation reaction: $CO_2(aq) + CO_3^{-2} + H_2O \rightarrow 2HCO_{3}^{-}$ (or, the reverse of disproportionation). Step 4\textrightarrow1 is concluded once the system returns to its equilibrium point at $A_i$ and $p_i$. Steps 3\textrightarrow4 and 4\textrightarrow1 could occur simultaneously, in principle, as could Steps 1\textrightarrow2 and 2\textrightarrow3.

\begin{figure}[h]
 \centering
 \includegraphics[width=\textwidth]{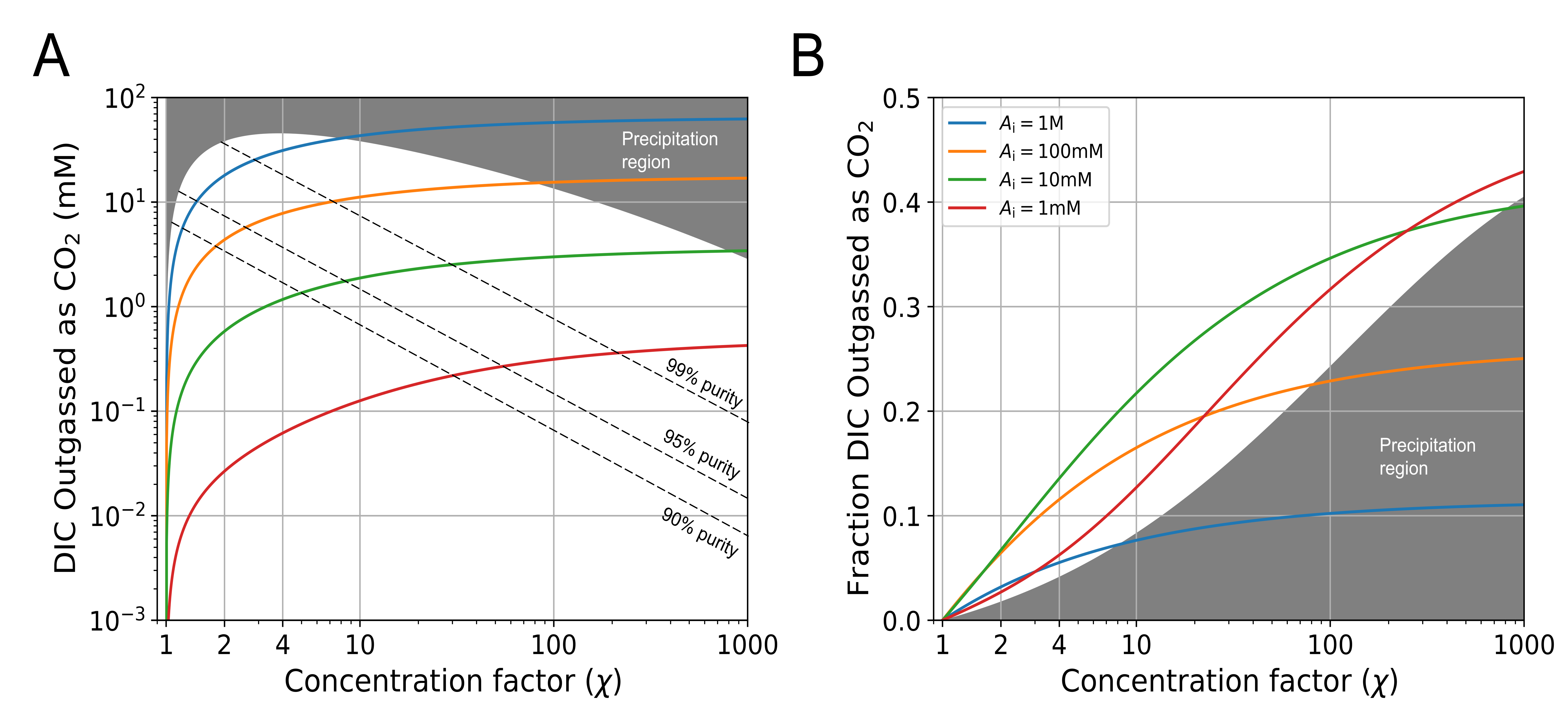}
 \caption{DIC outgassed based on the ACS. (A) Concentration of DIC in the feed solution outgassed as \ce{CO2} as a function of concentration factor. Dashed lines represent purity thresholds with respect to non-\ce{CO2} gases. (B) The fraction of DIC in the feed solution outgassed as \ce{CO2} as a function of concentration factor. Each color curve represents a different initial alkalinity concentration (legend in panel B applies also to panel A). The boundary of the gray region represents the precipitation threshold for potassium carbonate at 20°C of approximately 8M.}
 \label{fgr:acs_output}
\end{figure}

\subsection{\ce{CO2} Outgassed from the ACS}\label{sec:ACS_theory:outgas}

For a given temperature, the exact choices of parameters for the initial and final alkalinity values, as well as initial and final \ce{CO2} partial pressures, uniquely determine the outputs of the ACS. Whereas the initial pressure is set by the concentration of atmospheric \ce{CO2}, the outgassing pressure is a design parameter that should be set based on considerations relating to energy, rate, and water requirements. In this study, we fix the outgassing pressure at $p_f = 0.4$ mbar. The energetic considerations of the outgassing pressure are discussed briefly in Section \ref{sec:energy:extraction}, and will be the subject of future studies.

Figure \ref{fgr:acs_output} plots the result of the ACS for a fixed atmospheric and outgassing partial pressures of \ce{CO2} over a range of initial alkalinity values. The concentration factor specifies the concentration ($C_{out}$; Equation \ref{eq:outgas}) and fraction ($f_{out}$; Equation \ref{eq:fraction}) of DIC outgassed as \ce{CO2}. Outgassing purity thresholds are calculated based on partial pressures of other atmospheric gases (\ce{N2}, \ce{O2}, Ar; see Appendix Section B.4); higher concentration factors yield higher \ce{CO2} purity. In general, higher initial alkalinity values for the same concentration factor yield larger total outgassing values. The fraction of DIC outgassed exhibits a more complicated relationship with concentration factor. The lower the initial alkalinity, the higher the outgassed fraction can be (with an absolute limit at 0.5); the limiting regime is set when all DIC is in bicarbonate form and entirely disproportionates. As given by Equation \ref{eq:p_max}, the maximum outgassing pressure is set only by the concentration factor, invariant of the initial alkalinity. Increasing the concentration factor therefore increases the difference between the outgassing pressure ($p_f$) and the partial pressure of the solution in the concentrated state ($p_2$), which corresponds to higher absorption rates.\cite{zeebe_co2_2001} Table \ref{table:acs_values} lists output values for different representative ACS input parameters.

\begin{figure*}[h]
 \centering
 \includegraphics[width=0.9\textwidth]{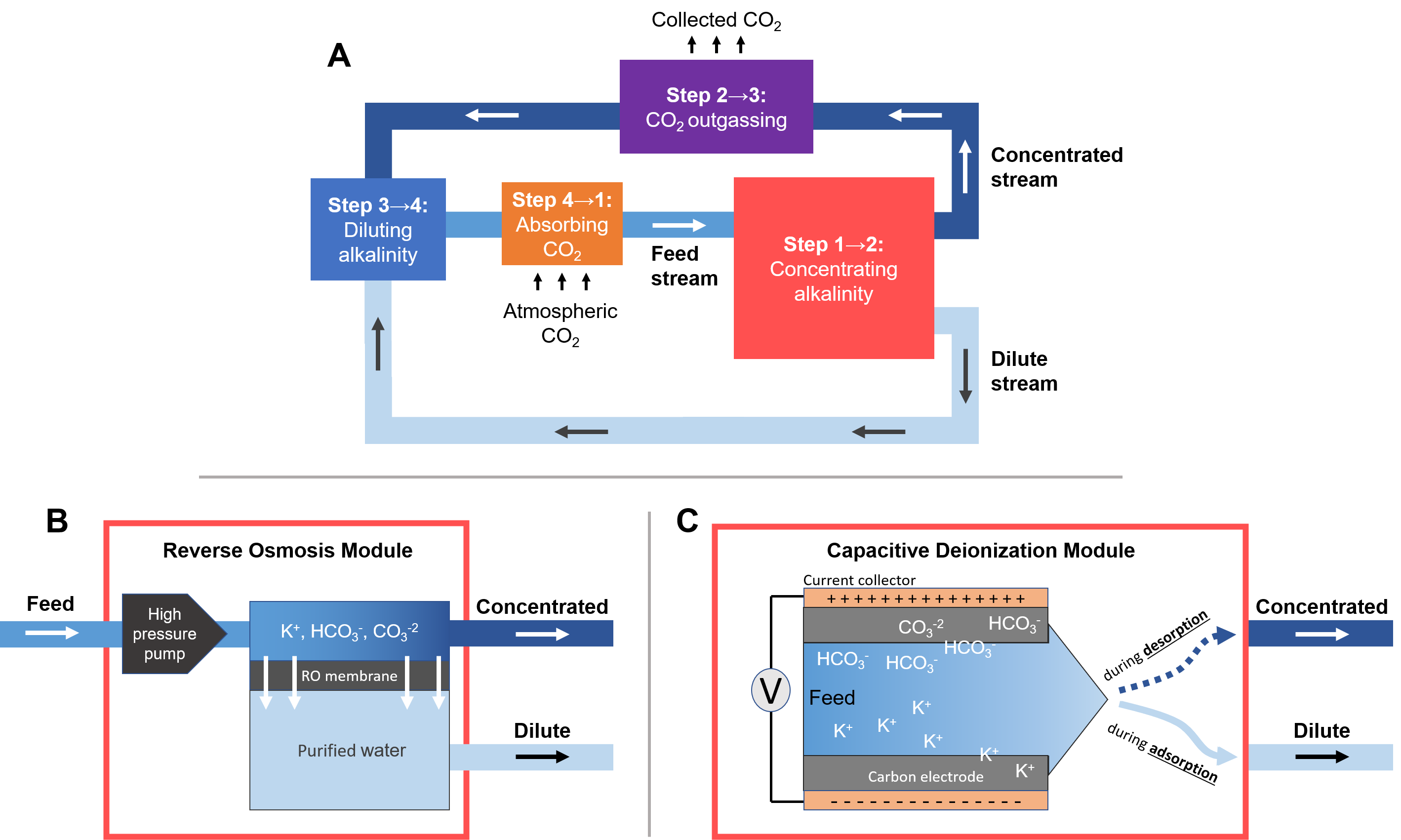}
 \caption{ACS system schematic. (A) The four steps of the ACS represented in a full cycle. (B) A schematic of a reverse osmosis module driven by a high-pressure pump. (C) A schematic of a capacitive deionization module driven by applied current and voltage. Ion exchange membranes are not represented in the diagram.}
 \label{fgr:schematic}
\end{figure*}

\section{Implementing the ACS}\label{sec:implementing_ACS}
The primary energy-consuming driving mechanism behind the ACS can be separated into two components: 1) a process to concentrate solutes in water, and 2) applying pressure for outgassing of \ce{CO2} from solution (Figure \ref{fgr:schematic}A). The remaining components, diluting alkalinity and absorbing \ce{CO2}, do not consume energy but are critical for evaluating water and contacting area requirements, as described in Section \ref{sec:comparison_and_scaleup}.

In principle, any desalination method, which produces purified water, can also be used to concentrate a stream of solute-filled solution. Desalination methods, for this reason, are candidate drivers for the ACS; they can be based on the following mechanisms: reverse osmosis (RO),\cite{fritzmann_state---art_2007,elimelech_future_2011} capacitive deionization (CDI),\cite{suss_water_2015} electrodialysis,\cite{al-amshawee_electrodialysis_2020} evaporation and distillation,\cite{alkhudhiri_membrane_2012,raj_review_2016} precipitation,\cite{shi_solar_2018} and solvent solubility.\cite{boo_membrane-less_2019} In this study, RO and CDI are considered for implementing the ACS. Both methods are deployed on an industrial scale, and evaluating the two serves as a comparison between pressure driven and electric field driven approaches.\cite{lin_energy_2019} Energetics of the ACS process using these two processes as examples are discussed in Section \ref{sec:energy}.

RO is a membrane-based separation process in which pressure is applied against a solvent-filled solution, overcoming the osmotic pressure of the solution, to create a concentrated and a dilute stream (Figure \ref{fgr:schematic}B). RO methods can be applied to brackish (low salinity) waters and wastewater processing with more dilute solutions, but are most commonly applied to seawater desalination. This application of RO is in broad commercial use, producing roughly 100 million cubic meters of purified water per day in 2018.\cite{jones_state_2019} Seawater desalination plants are typically designed to produce a stream of freshwater from an input feed of about 0.6 M of NaCl equivalent salt, yielding a concentrate output of roughly double the original salinity. In general, the RO process can be adapted to a broader range of initial salinities and higher overall concentration factors that may be desirable to achieve more optimal ACS output.

Existing technological developments and future prospects make RO an appealing candidate for implementing the ACS. For example, the development of energy recovery devices (ERDs) was crucial in reducing the power consumption of desalination to its current level. ERDs use the remaining energy stored in the pressure of the concentrate, which otherwise would be wasted, to apply part of the necessary pressure to the feed.\cite{fritzmann_state---art_2007} One significant difference between an ACS process and desalination is that in the ACS, after \ce{CO2} has been extracted, the concentrated and diluted solution are recombined. It is therefore possible to recover some of the energy held in the salinity gradient between the concentrated and dilute streams through forward osmosis (see Appendix Sections D.3.3 and D.3.4 for analytic treatment).\cite{straub_pressure-retarded_2016}

An alternative approach to the concentration step for ACS is CDI, which is a method of concentrating and removing anions and cations from solution by applying a voltage across two electrodes and creating an electric double layer made up of electrolyte ions (Figure \ref{fgr:schematic}C). When voltage is applied (<1.2 V to avoid splitting water), anions “electrosorb” to the positive electrode and cations to the negative electrode. When the voltage is switched off or reversed, the concentration of ions in the electrode pores and in the fluid between the electrodes sets the output concentration of a higher-alkalinity solution, driving the concentration step of the ACS. The material properties of the electrode (surface area, porosity, surface chemical groups, etc.) and the geometry constrain the overall capacity, rates, and energies of deionization.\cite{suss_water_2015} To increase efficiency, ion exchange membranes can be placed between the feed solution channel and the electrode material, in which case the process is called membrane CDI (MCDI).

Whatever approach is used to concentrate the alkaline solution, \ce{CO2} can be extracted from the concentrate stream by exposing it to a vacuum or a carrier gas. This can be done through a variety of standard methods in chemical engineering including vacuum pumps, or by making use of water vapor or another condensable gas (e.g. helium). Using a liquid-gas exchange membrane for \ce{CO2} extraction,\cite{willauer_recovery_2009,bhaumik_hollow_2004} which creates a gas-permeable barrier between gas and liquid phases, allows for lowering the outgassing pressure (<1 mbar) significantly below the vapor pressure of water while preventing flash evaporation.

Once \ce{CO2} is extracted from solution, the concentrated and dilute streams are combined, thereby diluting alkalinity to its initial concentration. At this point, the solution has less DIC relative to alkalinity than it would have at atmospheric conditions. Exposing this solution to the atmosphere will initiate an equilibration process of \ce{CO2} absorption, which is critical to evaluate in order to assess ACS requirements for water processing, water on hand, and land use. Due to the relatively low hydroxide ion concentrations associated with ACS conditions, air-liquid contactors, which increase the surface area of solution and use fans to increase exposure to air, are likely to be ineffective as a result of slow absorption kinetics. Instead, we can use large contacting reservoirs, potentially with mechanisms to enhance convective mixing. The kinetics of gaseous \ce{CO2} absorbing into water and reacting with hydroxide ions to form bicarbonate ions is the rate limiting step in the absorption process.\cite{stolaroff_carbon_2008} Overall absorption rate increases like the square root for higher hydroxide concentrations (i.e. higher pH) and linearly for higher air-liquid surface area. In Section \ref{sec:comparison_and_scaleup:water}, we use an approximated absorption rate to estimate the water on hand requirement for an ACS system given a certain facility water processing rate. The absorption time scale sets the duration that processed water needs to reside in the reservoir to reload DIC back into solution, and thus sets the total amount of water needed in the reservoir to operate the system in a continuous manner.

In the next section, we discuss the energy requirements of the ACS including the work associated with \ce{CO2} extraction. We introduce two simple energy models to explore ACS energy trade-offs, referencing a range of energy values based on reported values from RO and CDI systems.

\section{ACS thermodynamics and energy models}\label{sec:energy}

The process of concentrating ions in aqueous solution can be achieved, in general, by doing work to either confine ions to a smaller volume or to selectively remove water molecules from solution. Whereas the fundamental thermodynamic limit for the work required by a concentrating process is set by the entropic difference between the input and output streams, the particular mechanism for concentrating determines the additional associated dissipated energy. 

This section describes the thermodynamic minimum work of the ACS, discusses the associated vacuum outgassing energy requirements, and explores two high-level frameworks for evaluating the energetics of concentrating ions in solution to achieve Step 1\textrightarrow2 of the ACS. Two simplified energy models are proposed, one based on energy associated with reverse osmosis and another based on energy of binding ions in solution. Reverse osmosis and capacitive deionization are discussed as possible implementations of systems capable of concentrating ions to drive the ACS.

Irrespective of the particular concentrating mechanism, it is possible to set a thermodynamic limit on the ACS given an input and output partial pressure of \ce{CO2}. If the ACS cycle takes an input partial pressure of \ce{CO2}, $p_{i}$, and outgasses at a limiting output pressure of $p_{f,max}$, the thermodynamic minimum work per mole \ce{CO2} is given by: $w_{lim} = RT ln(p_{f,max}/p_{i})$, as long as it behaves as an ideal gas. Using carbonate chemistry assumptions (based on Equation \ref{eq:p_max}), we rewrite the thermodynamic minimum work expression in terms of the concentration factor, $\chi$, as: 

\begin{equation}
    w_{lim} = RT ln(\chi)
\label{eq:thermo_limit}
\end{equation}

If no vacuum is applied, a concentration factor of greater than 2500 — which is approximately equivalent to the ratio of non-\ce{CO2} molecules to \ce{CO2} molecules in the input stream of atmospheric air — is needed to outgas \ce{CO2} at 1 bar.

\subsection{Work needed for \ce{CO2} extraction from an aqueous solution}\label{sec:energy:extraction}
The nature of the ACS is such that the \ce{CO2} partial pressure limit in the concentrated state over the parameter range of interest (alkalinity between \ce{10^{-4}} to 10 M) is essentially proportional to the concentration factor (Equation \ref{eq:p_max}). In order to extract \ce{CO2} out of the concentrated solution, a \ce{CO2} partial pressure that is less than $p_{f,max}$ must be chosen. If the concentration factor is lower than $\sim$2500, a vacuum could be applied to the solution to extract \ce{CO2}, or the headspace could be filled with a separable carrier gas (e.g. helium or water vapor). For simplicity we analyze the application of vacuum. 

The thermodynamic minimum work needed to establish a vacuum at $p_f$ to permit \ce{CO2} outgassing isothermally is $w_{vac,min} = RT ln(p_{0}/p_{f})$, where $p_0 = 1$ bar; if $p_f =0.4$ mbar, then the minimum work is 19.1 kJ/mol\ce{CO2} (abbreviated as kJ/mol whenever referring to moles of \ce{CO2}). However, real physical systems incur additional losses from dissipation. Industrial vacuum pumps have process efficiencies in the range of 65-85\%.\cite{wilcox_carbon_2012}

There is a trade-off between the work needed to concentrate the feed solution and the work needed to establish a vacuum for extracting \ce{CO2} from solution. In this study, the work needed to establish the vacuum is held constant because we have chosen to fix the outgassing pressure, $p_f$, at 0.4 mbar, and subsequently bring the outgassed \ce{CO2} gas to 1 bar. Assuming an efficiency of $\eta = 70\%$ yields an additional work of $w_{vac} = w_{vac,min}/ \eta \approx 30$ kJ/mol. The trade-off that emerges from varying the outgassing pressure is a subject for future study.

Finally, because we assume the thermodynamic limit of isothermal compression, there is no additional required work to compress water vapor. Assuming it begins at its equilibrium vapor pressure at 20°C of 40 mbar, it precipitates during isothermal compression with no additional work, and its partial pressure remains at 40 mbar as the \ce{CO2} is compressed to a partial pressure of 1 bar.

\subsection{Reverse osmosis-driven ACS}\label{sec:energy:ro}
\subsubsection{Energy model}\label{sec:energy:ro:model}
We first pose a model for the work required to concentrate a solution through confining ions to a smaller volume. For an aqueous solution, when entropic effects are dominant, the relevant macroscopic state variables that determine free energy as the solution is concentrated and diluted are the osmotic pressure and concentration. The change in these state variables through the ACS sets the work necessary to concentrate the feed solution so that \ce{CO2} can be extracted.

In this “RO model,” we assume an idealized ion concentrating RO system in which water is driven through a perfectly selective semi-permeable membrane that blocks all non-water molecules. Given dilute conditions, the osmotic pressure ($\Pi$) across the membrane is determined by the Van’t Hoff approximation to be proportional to the solute concentration in the reference solution: $\Pi = RTC$. Here, $C$ refers to the sum of the total solute concentrations, accounting for anions, cations, and non-charged molecules (e.g, [\ce{K+}], [\ce{CO2]_{aq}}, [\ce{HCO3^-}], [\ce{CO3^-^2}]).

The following known effects are neglected in this model: interactions of ions in solution, entropic contribution based on the differentiation between solute species, partitioning of DIC species as the system is concentrated and diluted, and any membrane-specific effects, such as concentration polarization.\cite{fritzmann_state---art_2007,sablani_concentration_2001, qin_comparison_2019}\footnote[3]{RO processes account for the concentration polarization effect by running the maximum driving pressure roughly 10\% higher than the value set by the direct osmotic difference.} Indeed, a more complete model would account for the thermodynamic activities of species in the electrolyte mixture and would include a solution-diffusion component for membrane effects;\cite{baker_membrane_2004} such analyses may be the subject of future studies.

\begin{figure*}
 \centering
 \includegraphics[width=0.9\textwidth]{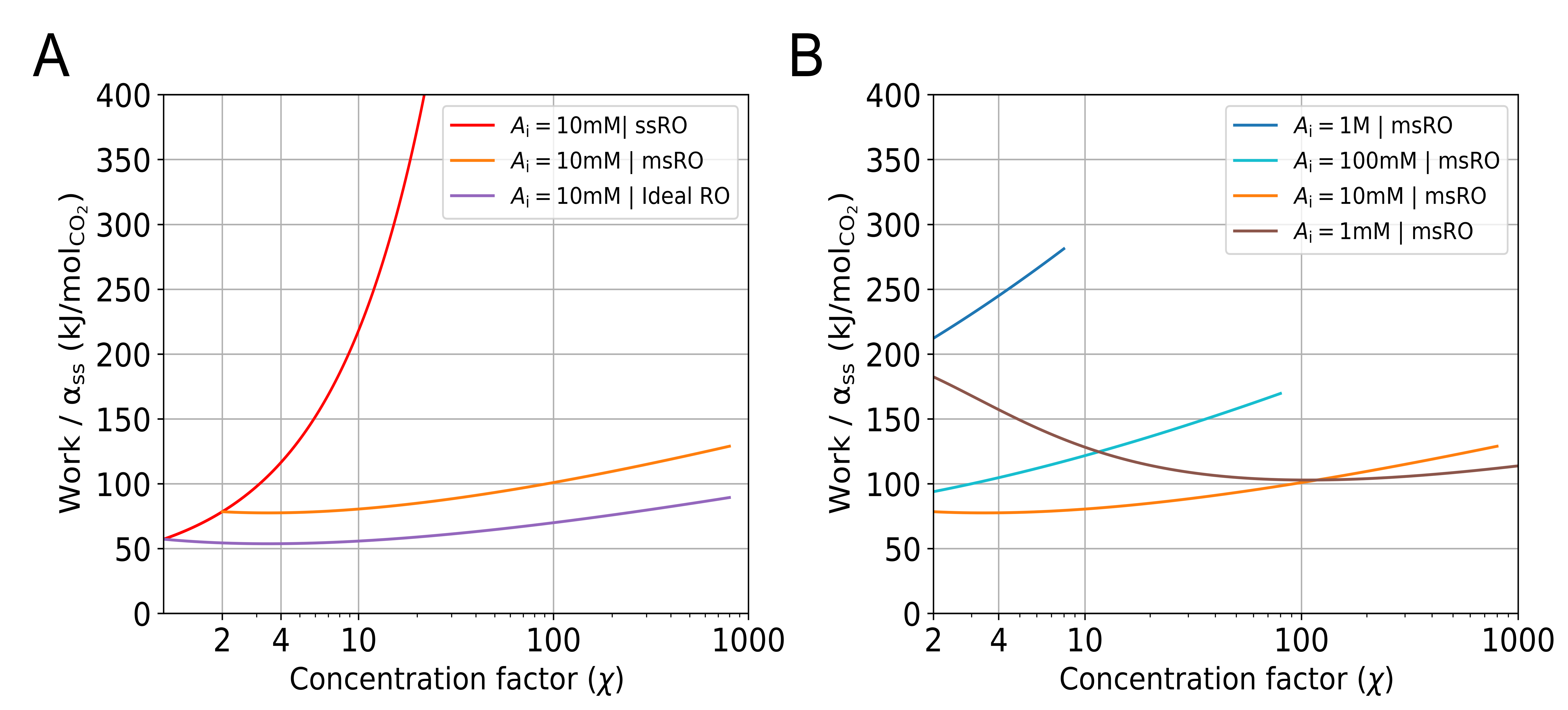}
 \caption{RO energy models. (A) Energy per mole \ce{CO2} as a function of concentration factor for three RO models: ideal, multi-stage, and single-stage at a representative initial alkalinity of 10 mM. (B) Energy of the multi-stage RO model evaluated at four different initial alkalinity values. (A-B): energy is normalized to the scaling parameter $\alpha_{ss}$; curves terminate on the right at 8M precipitation threshold for potassium carbonate; for all curves $p_f = 0.4$ mbar; msRO model assumes ssRO subcomponent of $\chi_{ss} = 2$. Associated vacuum work ($\sim$30 kJ/mol) is not included; see Table \ref{table:acs_values} for estimate of full implementation energies.}
 \label{fgr:ro}
\end{figure*}

The ideal RO work per mole of concentrated \ce{CO2} for a reversible process with no dissipation is:
\begin{equation}
    w_{RO,min} = RT (C_i/C_{out})ln(\chi)    
\end{equation}
Where $C_i = A_i + C_{DIC}(A_i, p_i)$, $\chi$ is the concentration factor, and $C_{out}$ is the concentration of outgassed \ce{CO2} calculated from carbonate equilibrium assumptions (see Appendix Sections D.3.1 and D.3.2 for full derivation of the RO model). The logarithmic scaling with $\chi$ requires significantly less work at lower initial alkalinities (see Figure \ref{fgr:ro}A). Physical systems approach this bound if the driving pressure is varied so as to be minimized throughout the entire concentration process.

A “single-stage” RO (ssRO) mode is driven by a single, fixed applied pressure throughout the concentrating process. Because we assume a perfectly selective membrane, the choice of the applied pressure is set only by the maximum concentration in the concentrated state. A single-stage system is simpler to construct but has higher energy dissipation because the applied pressure is substantially greater than the counteracting osmotic pressure in the early phase of the concentrating process. The work per mole in the ssRO process is given by: 
\begin{equation}
    w_{ssRO} = \alpha_{ss} RT(C_i/C_{out}) (\chi - 1)
\label{eq:ssRO}
\end{equation}
We include $\alpha_{ss}$ ($\geq 1$) as a scalable parameter to account for additional dissipation in physical RO systems; in the limiting case for ssRO, $\alpha_{ss} = 1$.

A “multi-stage” RO (msRO) process is made up of a series of ssRO modules. Instead of setting one driving pressure for the entire process, multiple driving pressures are chosen in order to reduce dissipation. If each ssRO subcomponent has an associated concentration factor of $\chi_{ss}$, the work per mole of \ce{CO2} is then: 
\begin{equation}
    w_{msRO} = \alpha_{ss} RT(C_i/C_{out}) (\chi_{ss} -1) log_{\chi_{ss}}(\chi)
\end{equation}
Here $\chi$ is still the overall concentration factor of the entire msRO system. We use the log scaling as a simplification, even though physical systems would typically be constructed from discrete single-stage modules and thus would more closely be expressed mathematically through a summation series (See Appendix Section D.3.5). 

In Figure \ref{fgr:ro}A, we compare the energy cost of the ssRO, msRO and ideal models at a single representative input alkalinity value (10 mM) over a range of concentration factors, normalized by the $\alpha_{ss}$ parameter. In the ssRO case, energy cost rapidly increases with concentration factor. In contrast, the msRO model, at the same initial alkalinity, is significantly more energetically favorable at high concentration factors than ssRO. The results of the msRO model are reported in Figure \ref{fgr:ro}B for various initial alkalinity values. The low initial alkalinity condition (1 mM) exhibits non-monotonic behavior, with a minimum around concentration factor of 100.

\subsubsection{Technological implementation}\label{sec:energy:ro:tech}
RO technology can be implemented to drive the ACS by applying pressure to selectively pass water through a semi-permeable membrane, thereby concentrating the remaining solution. RO systems can operate across a wide range of concentrations, but have been most technologically tailored for seawater conditions, and tend to be tuned around producing a low-concentration, potable solution, rather than optimizing the concentrate parameters.  

Typical seawater RO systems operate at around 80 bar and recover 50\% by volume of the saline feed (roughly 0.6 M of NaCl equivalent or 35 g/L) as freshwater. For the purposes of the ACS, concentrating 0.6 M of input alkalinity by a factor of 2 outgasses 13 mM of \ce{CO2} relative to the feed. We relate the energy cost of actual RO facilities to the theoretical ssRO work of 0.78 kWh per cubic meter of feed solution (typically reported as 1.56 kWh per cubic meter of freshwater), by evaluating the parameter $\alpha_{ss}$.\footnote[4]{The ideal RO work is 0.53 kWh per cubic meter of feed solution (or 1.06 kWh per cubic meter of freshwater).\cite{elimelech_future_2011}} Current energies of medium to large capacity industrial seawater systems usually range from 1.1 to 1.25 kWh per cubic meter of feed (i.e., $\alpha_{ss}$ ranging between approximately 1.4 and 1.6),\cite{penate_current_2012} with newer facilities regularly achieving lower than 1.0 kWh per cubic meter of feed ($\alpha_{ss} < 1.3$).\cite{elimelech_future_2011}

Brackish water RO tends to operate at lower salinities, typically 5-200 mM of NaCl equivalent. For the ACS at a concentration factor of 10, an initial alkalinity of 10 mM or 100 mM yields 1.9 mM or 11 mM, respectively, of extracted \ce{CO2} (Figure \ref{fgr:acs_output}A). Currently deployed brackish water systems tend to be less efficient than seawater RO systems, although they have the capacity of having a much lower required energy — below 1 kWh per cubic meter of feed.\cite{wilf_fundamentals_2014} For example, one industrial system that takes an input feed of 0.075 M NaCl equivalent with a concentration factor of 2 requires 0.445 kWh per cubic meter of feed ($\alpha_{ss} = 5.2$); another industrial system, which takes 0.063 M of NaCl equivalent feed with a concentration factor of 4 requires 0.825 kWh per cubic meter of feed ($\alpha_{ss} = 3.75$).\cite{zhao_energy_2013} Initial concentrations below 10 mM (such as 1 mM in Figure \ref{fgr:ro}) are not considered in this analysis because $\alpha_{ss}$ values are not reliably reported at such dilute conditions.

Detailed modelling studies have looked at how much industrial brackish water systems can be further optimized through the use of energy recovery devices and by tuning operating conditions. One simulation study reports 0.48 kWh per cubic meter of feed, given 0.25 M NaCl equivalent feed and concentration factor of 2.5 ($\alpha_{ss} = 1.1$).\cite{sarai_atab_operational_2016} Another detailed modelling study reports 0.30 kWh per cubic meter of feed, given 0.25 M NaCl equivalent feed and a concentration factor of 5 ($\alpha_{ss} = 1.2$).\cite{park_design_2020}

The significantly smaller values of $\alpha_{ss}$ from detailed modeling results compared to those for deployed industrial systems suggest that industrial brackish water systems can approach seawater systems in terms of energetic efficiency through straightforward modifications. The improvement in $\alpha_{ss}$ values is seen for systems with concentration factors ranging from about 2 to 5, suggesting that efficiency improvements could be applied to a wide range of system designs.

Although we rely on demonstrated RO systems to estimate the work required for driving the ACS, there are significant differences between ACS and desalination applications that must be considered. For example, brackish and seawater desalination must account for a complex variety of naturally occurring salt ion species and foulants present in seawater, which would not in general be the case in engineered ACS systems. On the other hand, RO membranes are not designed to be used for gas extraction and separating gases during the desalination process, which suggests further engineering modifications must be explored. Moreover, increased hydrostatic pressure as a result of the RO process modestly affects the carbonate equilibrium constants that are at the core of the ACS approach.

\subsection{Ion binding-driven ACS}\label{sec:energy:ib}
\subsubsection{Energy model}\label{sec:energy:ib:model}
Second, we pose a model for the work required to selectively remove water molecules from solution. This idealized “ion binding model" assumes that the energy to concentrate is dominated by enthalpic interactions, where a characteristic energy is associated with binding ions in solution, rather than entropic effects as assumed in the RO model. For a reversible ion adsorption process, the energy of binding ions from a feed solution and then releasing them into a concentrated stream sets the work necessary to concentrate the feed solution.

Here, we develop the simplest case consistent with experimental data:\cite{zhao_energy_2012} a constant electrical energy cost associated with binding an ion of a given charge out of the feed solution, independent of the concentration of ions found in the feed solution. We assume a value, $\epsilon_{ion}$, for the energy cost to bind a pair composed of a monovalent anion and a monovalent cation, and we double that for a pair of divalent ions ($2 \epsilon_{ion}$). This constant energy relationship may be observed when the selection mechanism applies charge or electric fields to do work on ions, rather than the uncharged water molecules of the solution, as in the RO energy model.

Such a model significantly simplifies physical effects as it neglects the following: ion-specific differences in binding energy, increasing binding energy as a function of number of bound ions, additional energy cost or energy recovered from unbinding the ion, including the concentration of the solution into which the ion is unbound, and entropic and electrostatic effects of confining ions to different concentrations.\cite{qin_comparison_2019} In general, entropic factors imply that the work to bind ions should depend at least weakly on solution concentration, and electrostatic factors imply that binding energy per ion will increase above some density of bound ions. Additionally, divalent ions may have different binding energies than pairs of monovalent ions, due to both entropic and enthalpic effects.

Nonetheless, the application of this formulation to the ACS allows us to evaluate the associated scaling relation of electrical work given initial and final alkalinities and partial pressures of \ce{CO2}. The result of the model is that the required work, per unit volume, to concentrate the feed stream, assuming monovalent cations and bicarbonate and carbonate anions, is then directly proportional to the concentration of ion charges in solution. We write the binding energy per mole of ions by setting $\epsilon_{m} = \epsilon_{ion} N_A$, where $N_A$ is Avogadro’s number. The charge concentration is then set by the alkalinity, so the total binding energy per volume and per mole is given by $\epsilon_{m} A_i$. The work per mole of outgassed \ce{CO2} is then (see Appendix Section D.4): 
\begin{equation}
    w_{IB} = \frac{\epsilon_{m} A_i}{C_{out}}
\label{eq:ib}
\end{equation}

\begin{figure*}
 \centering
 \includegraphics[width=\textwidth]{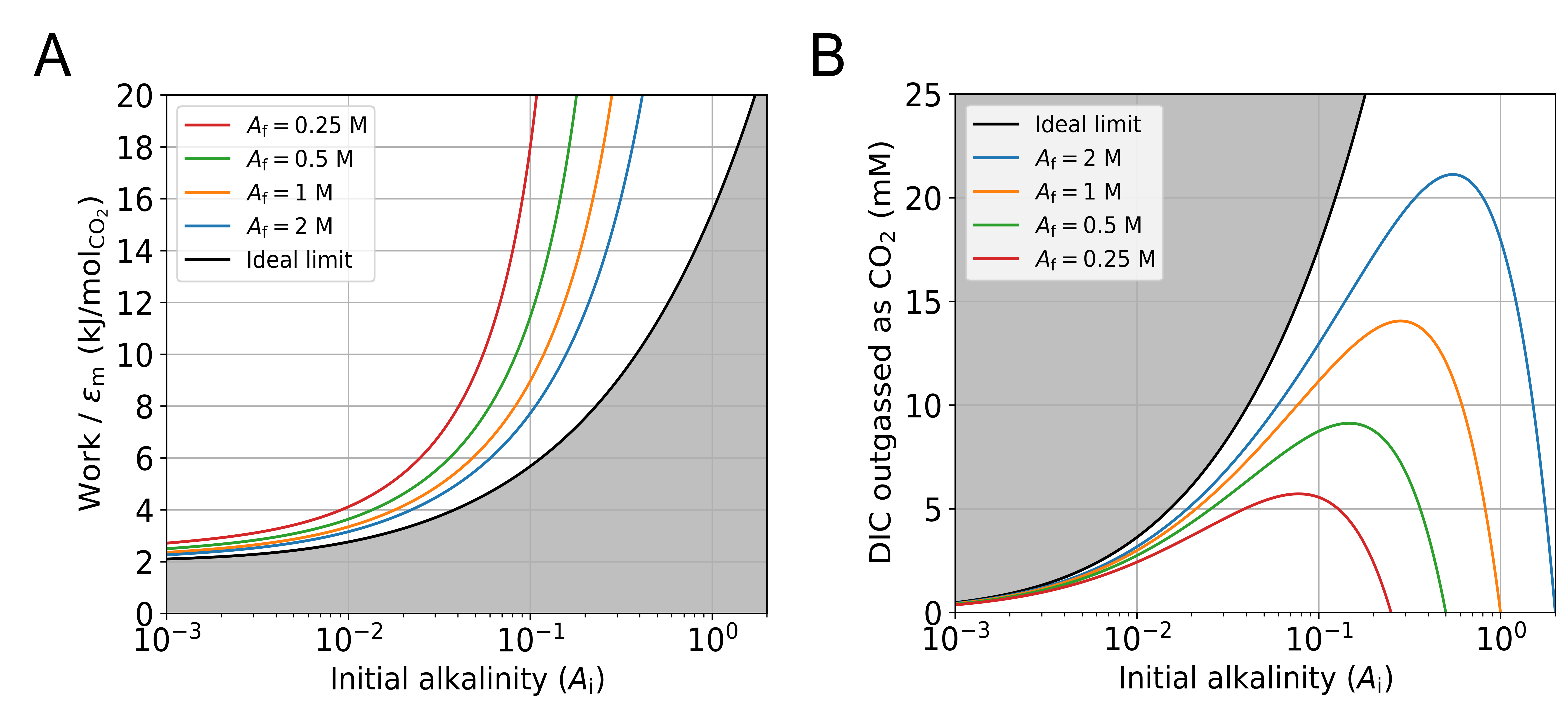}
     \caption{Results of ion binding model, calculated for various fixed initial alkalinity values. (A) Dependence on initial alkalinity of required work, normalized by $\epsilon_{m}$ to allow for rescaling to physical systems. Associated vacuum work is not included. In the ideal limit, the bicarbonate concentration fully disproportionates and half outgasses as \ce{CO2}. (B) Dependence on initial alkalinity of the concentration of DIC outgassed as \ce{CO2}. Curves cross 0 outgassed \ce{CO2} at the point where the initial alkalinity equals the final alkalinity. In all cases the final outgassing pressure is set to 0.4 mbar. The ideal limit here is the same as for A.}
 \label{fgr:ib}
\end{figure*}

In Figure \ref{fgr:ib}A shows the required work in the ion binding model per mole of \ce{CO2} vs. initial alkalinity for various final alkalinity values at an outgassing pressure of 0.4 mbar (a constant vacuum energy value must be added to compare the total necessary work). This type of plot is more useful to assess the ion binding model than plotting work vs. concentration factor, as in Figure \ref{fgr:ro}, because the physical and geometric properties of an ion-binding device are likely to set a constraint on the final alkalinity rather than the concentration factor. The minimum work per mole of outgassed \ce{CO2} is reached at the limit in which the feed stream of DIC consists entirely of bicarbonate ions at low alkalinity. The “ideal limit” indicates the limit in which, at high alkalinity, all of the bicarbonate ions disproportionate to carbonate ions and \ce{CO2} (Equation \ref{eq:disproportionation}) and a maximum of 50\% conversion is reached. At this limit the work is $2/\epsilon_{m}$ because two alkalinity carrier ions are bound for each \ce{CO2} molecule outgassed.

For any initial alkalinity, Figure \ref{fgr:ib} shows that the higher the final alkalinity, the higher is the $C_{DIC}$ outgassed as \ce{CO2} and the lower is the energy per mole of outgassed \ce{CO2}. In the context of this model, it is optimal to concentrate as much as physically possible as it does not penalize higher concentration factors. We note that, whereas energy efficiency is best for lower initial alkalinity values, the input stream $C_{DIC}$ is also accordingly low, which means more water handling is required per mole of outgassed \ce{CO2}.

We show in Figure \ref{fgr:ib}B that, for a given value of the final alkalinity, the total DIC outgassed as \ce{CO2} as a function of initial alkalinity exhibits a peak. This occurs because higher initial alkalinities hold higher DIC but, as the initial alkalinity approaches the final alkalinity value, a smaller fraction of that DIC is converted to \ce{CO2} and outgassed. This peak represents a further trade-off between outgassing concentration and outgassing energy built into the ACS as a result of the behavior of the carbonate system.

\subsubsection{Technology implementation}\label{sec:energy:ib:tech}
CDI technology can be implemented to drive the ACS by using electric fields to do work on ions at approximately a constant energy cost per ion. CDI systems tend to operate best in or just below brackish water salinities, with salt concentrations typically in the 5-200 mM range.\cite{suss_water_2015}

Specifically, Zhao \textit{et al.} showed experimentally that MCDI technology, which makes use of ion exchange membranes placed between the feed solution channel and the electrode, can operate at a value of $\epsilon_{ion}$ that is nearly independent of concentration. This occurs under constant current conditions over the entirety of the brackish water range, from 10-200 mM of NaCl.\cite{zhao_energy_2012} In this study, as in the ion binding model, this energy is also independent of the ion concentration in the concentrate stream. These results justify applying the ion binding model to ACS-CDI systems as a first-order study of energy scaling.

The values from MCDI studies are optimized given a condition on dilute stream purity, which, in the case of the ACS, unlike desalination, is not a relevant optimization target. In a limiting case, an excess of feed solution could be passed by the electrodes such that the dilute stream alkalinity is essentially the same as the feed. The potential to decrease $\epsilon_{ion}$ for differing solution optimization targets is an important subject for future studies.

Qin \textit{et al.} describe existing CDI systems that are capable, for a $\sim$30 mM salt feed stream, of recovering 95\% of the water while rejecting 90\% of the incoming salt,\cite{qin_comparison_2019} which corresponds to reaching a final alkalinity of $\sim$500 mM. Suss \textit{et al.} state that CDI can achieve 50\% water recovery for sea water,\cite{suss_water_2015} which would correspond to a final alkalinity of $\sim$1 M, and has the potential to achieve even higher water recovery ratios. This range of final alkalinities sets the regime explored in Figure \ref{fgr:ib}, with the understanding that existing CDI systems, designed for desalination and not necessarily a high water recovery ratio, have not yet been optimized to produce a high-alkalinity concentrate stream.

An additional property of capacitive systems, which depends on operating parameters such as cycle rate, flow rate, and the current and voltage control, is that some amount of energy is able to be recovered as ions are released back into solution and current is reversed. As described above, the ion binding model does not account for this energy recovered from unbinding an ion, but it can be added to the model as an overall energy recovery factor. Energy recovery values for experimental systems operating at optimal conditions have commonly been reported around 50\%,\cite{zhao_energy_2012,qin_comparison_2019,legrand_solvent-free_2018} with some studies approaching 80\%.\cite{dlugolecki_energy_2013}

\begin{table*}
\small
    \caption{ACS outgassing values and implementation energy estimates for various initial and concentrated alkalinities}
  \begin{tabular*}{\textwidth}{@{\extracolsep{\fill}}lll|lll|lll}
     \multicolumn{3}{c}{} & \multicolumn{3}{c}{ACS outgassing values} & \multicolumn{3}{c}{Energy estimates\footnotemark[9]}\\
    \hline
    &&&&&&&& MCDI  w/ 50\% \\
    $A_i$ (M) & $A_f$ (M) & $\chi$ & $C_{out}$ (mM) & $f_{out}$ &  $p_{f,max}$ (mbar) & RO (kJ/mol) & MCDI (kJ/mol) & Recovery (kJ/mol)\\
    \hline
    1e-2 & 1 & 100 & 3.0 & 0.34 & 40 & 160-190 & 310-730 & 170-380\\
    0.1 & 1 & 10 & 11 & 0.16 & 4.0 & 190-220 & 790-1900 & 410-970 \\
    0.6 & 1.2 & 2 & 13 & 0.039 & 0.80 & 250-310 & — & — \\
    1 & 4 & 4 & 31 & 0.055 & 1.6 & 350-420 & — & — \\
    \hline
  \end{tabular*}
  \caption*{$^{\ddagger\ddagger}$ {\footnotesize Here we assume $\alpha = 1.3 - 1.6$ and $\epsilon_{m} = 85-210$ kJ/mol, and we include the constant work to apply a vacuum at 0.4 mbar to be 30 kJ/mol. For the RO column, Row 3 corresponds to a ssRO model of $\chi = 2$ for approximate seawater conditions; all other rows are based on msRO using $\chi_{ss}=2$.}}
  \label{table:acs_values}
\end{table*}
  
\subsection{Comparing estimated ACS implementation energies}\label{sec:energy:comparison}
Using the simplified models for reverse osmosis-driven and ion binding-driven ACS, we use parameters from the literature for implementations of RO and CDI, respectively, to estimate the energies per mole for different ACS processes if implemented with real physical systems. Here, we seek to estimate the energy that would be required to complete the full ACS cycle and extract \ce{CO2} at 1 bar, so we include the additional work needed for vacuum outgassing, calculated in Section \ref{sec:energy:extraction}.

We caution that the values presented in this section are estimates that stem from plugging physical values into the simplified models above and do not represent a comprehensive prediction of the energy that will be required for these processes, at either industrial- or lab-scale. This will be the subject of future studies based on experimental work. The presentation of energies per mole, in Table \ref{table:acs_values}, is to provide a basis for understanding over which ACS parameter ranges either RO or CDI may be more practical and over which ranges each might be practical at all.

For ACS-RO, we recall that the energy values for current seawater RO systems tend to be between 1.0 to 1.25 kWh per cubic meter of feed, corresponding to an $\alpha_{ss}$ range between approximately 1.3 and 1.6. We plug this range into the plots in Figure \ref{fgr:ro} for an initial alkalinity of 0.6 M, corresponding to seawater, and a concentration factor of two, corresponding to typical seawater RO, to obtain the energy estimates for RO in Table \ref{table:acs_values}, Row 3.

In Rows 1 and 2 in Table \ref{table:acs_values}, representing brackish water alkalinities, $\alpha_{ss}$ ranges from 1.3 to 5.2, the latter coming from the higher end of the range of $\alpha_{ss}$ values for deployed brackish water systems. We justify extending the lower range of $\alpha_{ss}$ down to 1.3 for lower initial alkalinities because of the simulation studies, cited in Section \ref{sec:energy:ro:tech}, that suggest industrial-scale facilities could be built with even lower $\alpha_{ss}$ values. The feasibility of deploying ACS-RO for brackish water with an $\alpha_{ss}$ of 1.3 is further substantiated in analysis by Qin \textit{et al.} (2019), who show that the energy efficiency of existing brackish water RO deployments should not be taken as a limit because they have not yet reached the proper scale.\cite{qin_comparison_2019}

For ACS-CDI, we use the energies realized by MCDI systems because they can operate at a value of $\epsilon_{ion}$ that is nearly independent of concentration and are also uniformly more energetically efficient than CDI without ion exchange membranes operated under the same conditions. MCDI ion removal energy values range from approximately 17 to 42 kT per ion, or 85-210 kJ per mol NaCl equivalent salt.\cite{suss_water_2015} One experimental study reports a value independent of input concentration of 22 kT per ion, or 110 kJ per mol NaCl equivalent salt, for an MCDI system operating over the range 10-200 mM NaCl.\cite{zhao_energy_2012} We use Equation \ref{eq:ib} to obtain the energy estimates for MCDI in Table \ref{table:acs_values}.

As an example, we estimate the ACS-CDI energy for a particular set of parameters by applying the 110 kJ per mol value. If 10 mM input alkalinity is concentrated by a factor of 100 to 1 M, 3.0 mM of \ce{CO2} would outgas at 330 kJ/mol. If we are able to achieve the commonly reported 50\% energy recovery rate for the process, and then add the 30 kJ/mol for vacuum outgassing (Section \ref{sec:energy:extraction}), we would estimate the total work for the process to be 195 kJ/mol. That value falls in the range corresponding to these parameters in the Row 1 of Table \ref{table:acs_values}, under “MCDI w/ 50\% Recovery.”
\\
\section{Comparison to incumbent DAC technologies and feasibility of scale-up}\label{sec:comparison_and_scaleup}

In order for a particular DAC technology to contribute significantly to averting anthropogenic climate change, and contribute toward the gigatonne-per-year scale of global CDR that may be necessary by the end of the century, it needs to be able to be deployed at a large scale. To determine the feasibility of scale-up, we examine what challenges ACS implementations might face in reaching large-scale deployment levels and how these compare to incumbent technologies, using the benchmark set by the National Academies of Sciences (NAS) report evaluating 1 Mt\ce{CO2}/year DAC facilities.\cite{nasem_negative_2019} Whether through implementation using RO or CDI, the ACS possesses a number of potential advantages over incumbent DAC approaches but will also need to overcome some significant challenges.

\subsection{Energy}\label{sec:comparison_and_scaleup:energy}
The idealized energy requirement estimates provided for possible ACS-RO and ACS-CDI implementations in Section \ref{sec:energy}, while incomplete, can be compared to estimated requirements for incumbent DAC technologies. The lowest work from the conditions in Table \ref{table:acs_values} comes from Row 1, for which we estimate RO will require 160-190 kJ/mol and MCDI with recovery will require 170-380 kJ/mol. These estimates incorporate the work required to bring the outgassed \ce{CO2} gas to 1 bar. As shown in Appendix Section B.4, this condition attains a purity of 99.8\%. The NAS report estimates that the Carbon Engineering calcium loop-driven liquid solvent system has a work requirement of 360-480 kJ/mol (reported as 8.2-11 GJ/t) and that solid sorbent systems, for more realistic “mid-range scenarios,” have an energy requirement of 174-261 kJ/mol (reported as 3.95-5.92 GJ/t).\cite{nasem_negative_2019} These ranges are both for systems operating at a scale of 1 Mt\ce{CO2}/year removed and for conditions comparable to our assumptions, capturing from a 400 ppm atmosphere at 25°C, with a 98\% purity product.\footnote[7]{NAS report energy ranges are based on a \ce{CO2} capture efficiency of 75\%. We do not directly consider capture efficiency for the ACS, however, because ACS processes rely on passive contacting pools. Instead, as described in Section \ref{sec:comparison_and_scaleup:water}, we consider the timescale associated with the passive equilibration of those pools.}

While the ACS idealized energy requirement ranges fall below that of the liquid solvent system and within the approximate range of solid sorbent systems, these ACS values should be viewed as far more uncertain when compared to ranges based on systems that have been realized at demonstration scale.\footnote[8]{For example, the energy of liquid pumping has been neglected in this analysis. If the scale of the additional energy cost per mole of \ce{CO2} due to pumping is roughly approximated by the work needed to raise all processed water by 10 m then, for the conditions for which 3.0 mM and 31 mM of \ce{CO2} is outgassed (Table \ref{table:acs_values}), this would correspond to roughly an additional 30 and 3 kJ/mol, respectively.} We note that the work estimate for MCDI can be improved if higher energy recovery factors, approaching the 80\% factor reported for some systems,\cite{dlugolecki_energy_2013} can be attained, though MCDI quickly becomes unfavorable at higher initial alkalinities when compared to both ACS-RO work requirements and to incumbent technologies. For ACS-RO, the work requirement increases less quickly for higher initial alkalinities and each of the other conditions in Table \ref{table:acs_values} remains in the range of the liquid solvent system.

Even without a final accounting of the energy requirements for implementing the ACS, we can compare the sources of energy needed to incumbent DAC technologies. In the Carbon Engineering process, the core calcining step, for the final release of concentrated \ce{CO2} from calcium carbonate precipitate, requires heating to $\sim$900°C.\cite{nasem_negative_2019} Even though heat recovery is used for other processes that require low-grade heat and electrical energy is chosen when it can be used efficiently, the Carbon Engineering process still uses 5.25 GJ of natural gas per tonne of captured \ce{CO2}.\cite{keith_process_2018} Though the cost and energy requirements of the Carbon Engineering process account for offsetting of direct emissions, scaling up such a process would entrench an ongoing need for production of natural gas, as well as any emissions associated with the natural gas supply chain.\cite{zhang_quantifying_2020} Incumbent technologies relying on solid sorbents typically require heating to $\sim$100°C during the desorption step of a thermal swing.

In contrast, the ACS does not require heat to operate, and can be operated entirely through renewable energy sources. Outside of the module that operates the ACS cycle itself, fluid pumping, for vacuum extraction and solvent pumping, can all be powered by electrical energy. An ACS-CDI module would require only electrical energy for ion binding,\cite{suss_water_2015} as would an ACS-RO module operating in the majority of the optimal regime described in Section \ref{sec:energy}.\cite{fritzmann_state---art_2007}

\subsection{Water}\label{sec:comparison_and_scaleup:water}
For any DAC approach that makes use of a liquid solvent for the initial capture of carbon dioxide from the air, it is important to determine the feasibility of the requirements for both overall water volume processing and water on hand. The ACS, as shown in Table \ref{table:acs_values}, can use an incoming solution ranging from 10 mM to 1 M alkalinity, resulting in respective output concentrations ranging from 3.0 mM to 31 mM.

For 10 mM initial alkalinity solution, removing 3.0 mM each cycle means \SI{7.6e9}{m^3} of water volume needs to be processed to remove 1 Mt\ce{CO2} total. This is roughly an order of magnitude more water than the annual processing rate for a large RO facility today, as we describe in Section \ref{sec:comparison_and_scaleup:feasibility}. For 1 M initial alkalinity solution, removing 31 mM each cycle means \SI{7.4e8}{m^3} of water needs to be processed, decreasing the processing requirement to roughly that of a large RO facility.

The liquid solvent system of Carbon Engineering is the key incumbent DAC technology that has significant water use. Carbon Engineering inputs 35,000 tonnes of a 0.45 M \ce{CO3^-^2} and 2.0 M \ce{K+} solution into its contactor per hour and uses it to capture 112 tonnes of \ce{CO2} per hour,\cite{keith_process_2018} for a captured \ce{CO2} concentration of 73 mM. Per unit of \ce{CO2} removed, an ACS system would then require between 2.4 and 24 times as much water to be moved through the system each cycle.

The Carbon Engineering system requires an incoming solution stream of high alkalinity and high DIC. The ACS, however, takes a dilute incoming solution stream that is less constrained to a particular concentration. Air contact can thus be achieved with the additional surface area of the dilute solution, for example, with large pools for passive contacting. Large pools, with a high surface-area-to-volume ratio have significant water losses to evaporation as a function of humidity — these losses can be mitigated by either locating facilities in humid or rainy regions or by replenishing the pools. A dilute incoming solution stream also requires an increased energetic cost for liquid handling, although it reduces the energy requirement for operating contacting fans.

An ACS facility would require significant amounts of water to be stored on hand in reservoirs near the facility due to the slow equilibration of the \ce{CO2}-depleted solvent stream as it absorbs \ce{CO2} before again being cycled through. These pools are where the feed stream for the ACS process would be drawn and where the combined concentrated and dilute streams would be returned, as shown in Figure \ref{fgr:schematic}A. For a 1 Mt\ce{CO2}/year facility, assuming large passive contacting pools of 0.1 meter depth, Stolaroff \textit{et al.} provide a method for estimating the passive \ce{CO2} uptake rate as a function of alkalinity.\cite{stolaroff_carbon_2008, danckwerts_absorption_1950} Using the ACS conditions, an instantaneous uptake rate of \SI{5e-7}{mol/s/m^2} of area for a 10 mM solution and \SI{8e-7}{mol/s/m^2} for a 1 M solution. If we use this as an estimate for the rate throughout the equilibration process, for the 3.0 mM and 31 mM \ce{CO2} extraction quantities in Table \ref{table:acs_values} we obtain a characteristic equilibration timescale, $\tau$, of 7 days and 50 days, respectively. For the 10 mM condition, this would mean keeping roughly $10^8 \text{ m}^3$ of water on hand in the pools, equivalent to the volume of approximately 40,000 Olympic swimming pools, cycled approximately 50 times per year.

These rates are applicable for the concentration of hydroxide ions in the maximally \ce{CO2}-depleted solution, however, and therefore represent the largest absorption rate over the duration of the equilibration process, resulting in a lower bound for the equilibration timescale. Assuming that this uptake rate slows approximately exponentially, for an elapsed time $t$ we expect only the first $1 - e^{-t/\tau}$ of the equilibration process to have been completed. Over 7 days we would expect approximately 60\% of the equilibration process to have completed. Cycling this solution through an ACS system would result in approximately 40\% less \ce{CO2} to be extracted for a similar energy cost. An engineering tradeoff between rate and energy cost would thus need optimization; e.g. waiting twice as long permits the equilibration to proceed to roughly 90\% completion but requires keeping roughly \SI{3e8}{m^3} of water on hand.

There is also a variety of ways to mitigate the extended timescales estimated above by introducing mixing into the contacting pools, instead of equilibrating in passive, unmixed pools. Depending on area constraints and the type of mixing used, this could be done either by continuously combining the outlet streams from the ACS system into one contacting pool, achieving a steady-state DIC concentration, or by using many smaller pools. Mixing can be done actively, as is done at some water treatment facilities,\cite{cong_new_2009} at an additional energetic cost. If we choose locations with high enough ground wind speeds, we can also expect the pools to remain well-mixed by the wind.\cite{kullenberg_vertical_1976}

\subsection{Land}\label{sec:comparison_and_scaleup:land}
As land continues to become a more limited resource, it is important to determine the land use requirement for any DAC technology. This land use requirement includes the footprint of the DAC facility itself (which, for the ACS, is predominated by the footprint of passive contacting pools) and the footprint of generation facilities required to power the DAC facility.

We estimate the land area requirement to power an ACS facility using only renewable energy sources, which requires substantially more land than using fossil energy, by using a result from Fthenakis and Kim that 1 Mha of land is required for 78.2 GW of solar capacity.\cite{fthenakis_land_2009} Taking the upper end of the range for ACS-RO for a 10 mM solution, 190 kJ/mol (or 4.3 GJ/t\ce{CO2}), we then determine that we would need \SI{1.7e3} ha for the solar power to produce 1 Mt\ce{CO2}/year. This land use requirement for power generation is much smaller than that for the contacting pools. From the estimate above in Section \ref{sec:comparison_and_scaleup:water}, to remove 1 Mt\ce{CO2}/year with pools of 0.1 meter depth would require $10^9 \text{ m}^2$ ($10^5$ ha). For context, this is a large area — the Great Salt Lake is about four times the size.

We compare the ACS to incumbent technologies’ land use requirement only for the facility itself, as the power generation footprint for each technology simply scales with the amount of power needed for a particular type of power source. Even when accounting for the indirect land impact, to ensure there are no detrimental environmental impacts when multiple facilities are built in the same area, the Carbon Engineering approach requires only approximately 700 ha to produce 1 Mt\ce{CO2}/year, and much of this land could likely be used more flexibly; solid sorbent approaches require even less land.\cite{nasem_negative_2019}

To scale up the ACS to gigatonne-scale using passive contacting pools would require a likely unfeasibly large portion of land, on the order of 10\% of the area of the US. Implementation of active mixing approaches, described in Section \ref{sec:comparison_and_scaleup:water}, could reduce this land requirement, though would carry an energetic cost. If mixing is used, the land use requirement can be reduced by an even larger factor than the water on hand requirement because, without the need for passive contacting, much deeper pools could be used.

\subsection{Feasibility of scale-up and opportunities for optimization}\label{sec:comparison_and_scaleup:feasibility}
An important advantage of the ACS approach to DAC is its ability to leverage existing technologies for water purification and desalination that are widely deployed at commercial scale around the world. Large RO facilities have capacities of more than 50 million m\textsuperscript{3} of purified water per year, with the largest plants having capacities of more than 350 million m\textsuperscript{3} of purified water per year.\cite{jones_state_2019} As described in Section \ref{sec:comparison_and_scaleup:water}, plants of this largest size implementing ACS (for example, using the condition from Table \ref{table:acs_values}, Row 4 outgassing 31 mM of \ce{CO2}) would be able to capture up to 1 Mt\ce{CO2}/year. Global desalination capacity is currently roughly 35 billion m\textsuperscript{3} of water per year and growing rapidly. So, achieving a scale on the order of 100 Mt\ce{CO2} captured per year seems feasible based on current RO deployments, with larger scales achievable over time, though this would likely require substantial improvements in land use and water on hand requirements. 

From the preceding sub-sections, we observe a clear trade-off between the required energy use, water on hand, and land use and the total water volume processing capacity to deploy a 1 Mt\ce{CO2}/year ACS-DAC facility. In Table \ref{table:acs_values}, we see the general trend that higher initial alkalinities have increasing energy requirements. On the one hand, whereas the 1 M initial alkalinity condition in Row 4 remains in the range of the incumbent liquid solvent DAC system for RO, this condition is far too energetically costly for CDI. On the other hand, we showed in Section \ref{sec:comparison_and_scaleup:water} that the 10 mM initial alkalinity condition in Row 1 requires roughly an order of magnitude more water to be processed than the Row 4 condition, which itself already requires roughly the water processing of a large RO facility to achieve 1 Mt\ce{CO2}/year. Without any improvements, then, the water processing scale for this low alkalinity Row 1 condition may be very difficult to achieve. Because its equilibration rate is more than twice as fast as the Row 2 condition, however, the low alkalinity Row 1 condition is favored with significantly lower water on hand and land use requirements. These trade-offs demonstrate both the challenges in scaling ACS-DAC for most conditions and the possibilities for tailoring the optimal ACS implementation.

Although contacting requirements pose challenging land and water demands, several mitigation options are possible to increase absorption kinetics and outgassing efficiency. Beyond the simple conditions examined in this study, different choices of solvents and membranes could enhance the ACS. Whereas only a strong base solvent was considered here, preliminary analysis shows that the right choice of a weak base solvent could increase the ACS outgassing amount (See Appendix Section C). In CDI systems the use of an ion-exchange membrane tuned to select for bicarbonate ions over carbonate ions could increase ACS efficiency. More generally, engineering membrane or electrode properties around bicarbonate and carbonate ions is an important area of future study. Beyond the two specific technologies explored in this analysis, the broad suite of existing desalination approaches,\cite{lin_energy_2019} as well as hybrid approaches that combine strengths of different methods,\cite{ahmed_hybrid_2020} could be investigated as driving mechanisms for the ACS. Furthermore, principles from the ACS could be explored as a way of modifying and enhancing other solvent-based DAC methods. Finally, as described in Section \ref{sec:comparison_and_scaleup:water}, there are several ways to mix the contacting pools to increase the liquid-air contact area and increase the \ce{CO2} absorption rate.

In addition to ensuring equitable allocation of scarce land and water resources, associated environmental hazards and material considerations must be considered before ACS systems are scaled up. Environmental impacts from traditional desalination facilities come predominantly from discharge of brine and from chemical treatment of water and membranes. Current membrane cleaning methods make use of toxic substances, which would need to be disposed of safely if large-scale systems are deployed.\cite{tularam_environmental_2007} Because ACS systems can be operated in a closed cycle and, once operational, no significant discharge or uptake is necessary, these environmental harms can likely be mitigated.\cite{fritzmann_state---art_2007} Although no significant loss of potassium or any other alkalinity carrier is expected, scaling up large alkaline reservoirs would need to be accompanied with local environmental assessments to evaluate risk and determine mitigation options for spills or leakage.

\section{Conclusion}\label{sec:conclusion}
The Alkalinity Concentration Swing is a new approach to DAC in which the driving mechanism is based on concentrating an alkaline solution that has absorbed atmospheric \ce{CO2}. Concentrating a solution with a given alkalinity and DIC results in disproportionation of bicarbonate ions into aqueous \ce{CO2} and carbonate ions, proportionally increasing the outgassing partial pressure (Equation \ref{eq:p_max}). This allows for extraction and compression of \ce{CO2}. For the same concentration factor, higher initial alkalinity solutions outgas a greater amount of \ce{CO2} relative to the initial feed (Figure \ref{fgr:acs_output}B). For a given final alkalinity, the amount of \ce{CO2} outgassed vs. initial alkalinity exhibits a peak as initial alkalinity approaches that final alkalinity because of a trade-off between higher DIC available and a smaller conversion fraction of that DIC (Figure 5B).

The ACS can be implemented based on desalination technologies. We propose and briefly evaluate two technological implementation approaches, RO and CDI, with two accompanying simplified energy models, the RO model and the ion binding model, respectively. For each, the \ce{CO2} capture energy (Table \ref{table:acs_values}) is dependent on the initial alkalinity, the concentration factor, and the applied vacuum pressure, as well as the dissipation for the associated implementation mechanism. We use reported experimental values from existing RO and CDI desalination implementations to estimate associated capture energies for ACS-RO and ACS-CDI. The choice of initial alkalinity and concentration factor present trade-offs between the quantity of outgassed DIC and energy requirements (Table \ref{table:acs_values}). For example, a solution initialized at 10 mM alkalinity and concentrated by a factor of 100 outgasses approximately 3 mM of \ce{CO2} and requires a lower bound of 160 and 170 kJ/mol for the msRO and MCDI (with energy recovery) models, respectively. A solution that swings between 1 M and 4 M alkalinity, however, outgasses 31 mM of \ce{CO2}, which corresponds to a factor of 10 less water processing, but requires a factor of about two more energy (lower bound of 350 kJ/mol, given the msRO model). Whereas the ACS idealized energy requirement ranges fall below that of the liquid solvent system and within the approximate range of solid sorbent systems, these ACS values should be viewed as far more uncertain when compared to ranges based on systems that have been realized at demonstration scale.

Overall, the ACS appears to be an intriguing DAC method in need of experimental research to further evaluate its viability. In particular, studies of ACS kinetics are important in order to understand possible limitations in the absorption and outgassing steps. Our analysis further reveals a trade-off for the ACS between the total water processing requirement and both the capture energy demand and water on hand requirement. Although initial calculations point to challenging land and water requirements to scale up this technology, we propose several potential mitigation pathways. Both ACS-RO and ACS-CDI approaches can be run entirely on electricity and do not rely on heat; the required materials are relatively simple (\ce{K+}, membranes, electrodes, water). An initial assessment points to relatively environmentally safe deployment because no toxic chemicals, such as amines, are critical for this process. Lastly, the proposed approaches are based on existing technologies that have been deployed at large scale, and significant research and development can be leveraged from the desalination industry.

\section*{Conflicts of interest}
There are no conflicts to declare.

\section*{Acknowledgements}
This research was supported by a grant from the Harvard University Climate Change Solutions Fund. We thank David Kwabi, David Keith, Jen Wilcox, Martin Jin, and Eric Fell for helpful discussions.

\newpage

\begin{appendices}
\counterwithin{figure}{section}
\counterwithin{table}{section}
\counterwithin{equation}{section}
\renewcommand\thefigure{\thesection\arabic{figure}}
\renewcommand\theequation{\thesection\arabic{equation}}
\renewcommand\thetable{\thesection\arabic{table}}
\addcontentsline{toc}{section}{Appendix}

\begin{center}
\textbf{\LARGE Appendix: Alkalinity Concentration Swing for Direct Air Capture of Carbon Dioxide} 
\end{center}

\section{Dissolved Inorganic Carbon Equilibrium}
\subsection{Carbonate equilibrium relations}
Dissolved inorganic carbon concentration ($C_\text{DIC}$), is defined as the sum of the concentration of the following three molecular species dissolved in aqueous solution:
\begin{equation}
C_\text{DIC} \equiv \text{[CO$_2]_\text{aq}$ + [HCO$_3^{-}$] + [CO$_3^{-2}$]}
\end{equation}
The equilibrium balance between the species is given by the following chemical reactions:
\begin{equation}
    \text{CO$_2$(gas) $\xleftrightarrow{H^\text{cp}}$ CO$_2$(aq)}
\end{equation}
\begin{equation}
   \text{CO$_2$(aq) + H$_2$O $\xleftrightarrow{K_1}$ H$^+$ + HCO$_{3}^{-}$ $\xleftrightarrow{K_2}$ 2H$^+$ + CO$_3^{-2}$}
\end{equation}
The equilibrium relations are:
\begin{equation}
    \text{[CO$_2]_\text{aq}$} = H^\text{cp}p_\text{CO$_2$}
\end{equation}
\begin{equation}
    K_1 = \frac{\text{[H$^+$][HCO$_3^-$]}}{\text{[CO$_2]_\text{aq}$}}
\end{equation}
\begin{equation}
    K_2 = \frac{\text{[H$^+$][CO$_3^{-2}$]}}{\text{[HCO$_3^-$]}}
\end{equation}
\begin{equation}
    K_\text{w} = \text{[H$^+$][OH$^-$]}
\end{equation}

H$_2$CO$_3$, or undissociated carbonic acid, appears in low concentrations relative to the other species in equilibrium, so it can be neglected. Its presence, however, may be important for evaluating kinetics In general, the bicarbonate and carbonate equilibrium relations, $K_1$ and $K_2$, depend on temperature, ionic strength (or salinity), and the species composition of the solution,\footnote{Roy, R.N., et al., The dissociation constants of carbonic acid in seawater at salinities 5 to 45 and temperatures 0 to 45 C. Marine Chemistry, 1993. 44: p. 249-267.} which is not evaluated in this analysis for simplicity. Throughout this study, we assume a solution at normal temperature (20$^{\circ}$ C) and pressure (1 atm) of zero salinity, $K_1 =$ \SI{9.6e-7}{M} and $K_2 =$ \SI{3.4e-10}{M}, resulting in the first and second $pK_a$ for the carbonate system being 6.0 and 9.5, respectively. Henry's coefficient is $H^\text{cp} =$ \SI{0.034}{M/bar}, and the dissociation constant of water is $K_\text{w} =$ \SI{1e-14}{M^2}. We take the partial pressure of \ce{CO2} ($p_\text{CO$_2$}$) to be 0.4 mbar (or 400 ppm, which is approximately atmospheric partial pressure of \ce{CO2}).

\subsection{DIC as a function of alkalinity and partial pressure of \ce{CO2}} \label{fixed_pCO2}

In equilibrium, at a given temperature and total pressure, defining both the alkalinity of solution and the partial pressure of \ce{CO2} uniquely specifies the state of the system with respect to total DIC concentration, relative concentration of carbon species, and pH.

In this analysis, alkalinity is defined as the molar charge difference between the sum of the positive and negative conservative ions in solution.\footnote{Zeebe, Richard, and Dieter Wolf-Gladrow. CO2 in Seawater: Equilibrium, Kinetics, Isotopes, 2001.} Here “conservative” refers to ions whose concentration in an aqueous environment is insensitive to pH, pressure, or temperature. While in the simplified system we consider here we assume a monovalent positive charge alkalinity carrier (specifically, K$^+$), in general, alkalinity ($A$) is defined in the following way:
\begin{equation}
\begin{split}
    A \equiv \; & [\text{Na}^+] + [\text{K}^+] + 2[\text{Mg}^{2+}] + 2[\text{Ca}^{2+}] + (\mbox{other conservative cation charge}) \\
    - & [\text{Cl}^-] - [\text{Br}^-] - (\mbox{other conservative anion charge})
\end{split}
\end{equation}
The factor of 2 on the magnesium and calcium ion concentrations is due to the divalent charge on those ions. The relationship between alkalinity and all non-conservative ions, including DIC species, derives directly from charge conservation. The charge-neutrality condition requires that the excess charge of conservative cations over conservative anions equal the excess charge of non-conservative anions over non-conservative cations, represented in our system in the following way:
\begin{equation}
    A = \text{[HCO$_3^-$] + 2 [CO$_3^{-2}$] + [OH$^-$] - [H$^+$]}
\end{equation}

Next, we write down the entire system of equation that relates all the DIC species to alkalinity and $p_\text{CO$_2$}$. For readability, we simplify the chemical species names to variables: $h \equiv$ [H$^+$]; $a \equiv $ [CO$_2]_\text{aq}$; $b \equiv$ [HCO$_3^{-}]$; $c \equiv$ [CO$_3^{-2}]$. The water disassociation constant $K_\text{w}$ allows us to write [OH$^-] = K_\text{w}/h$. This gives the following system of equations:
\begin{equation}
    K_1 = \dfrac{hb}{a_0}
    \label{eq:K1}
\end{equation}
\begin{equation}
    K_2 = \dfrac{hc}{b}
    \label{eq:K2}
\end{equation}
The fixed partial pressure condition sets an unchanging equilibrium between gaseous and aqueous \ce{CO2}, such that $a_0 = H^\text{cp} p_\text{CO$_2$}$. For $p_\text{CO$_2$} = 0.4$ mbar, $a_0 \approx$ \SI{1.4e-5}{M}.

The alkalinity charge-balance equation is then:
\begin{equation}
    A = b + 2c + K_\text{w}/h - h
    \label{eq:alkalinity}
\end{equation}

For a fixed A, and writing the expressions for $b$ and $c$ as function of $h$, we use a computational root solver to find a value for h given the following relationship:
\begin{equation}
    A = H^\text{cp} p_\text{CO$_2$} \left( \dfrac{K_1}{h} + \dfrac{2 K_1 K_2}{h^2} \right) + K_\text{w}/h - h
    \label{eq:alk_pco2}
\end{equation}

\begin{figure}[h]
    \centering
    \includegraphics[width=10cm]{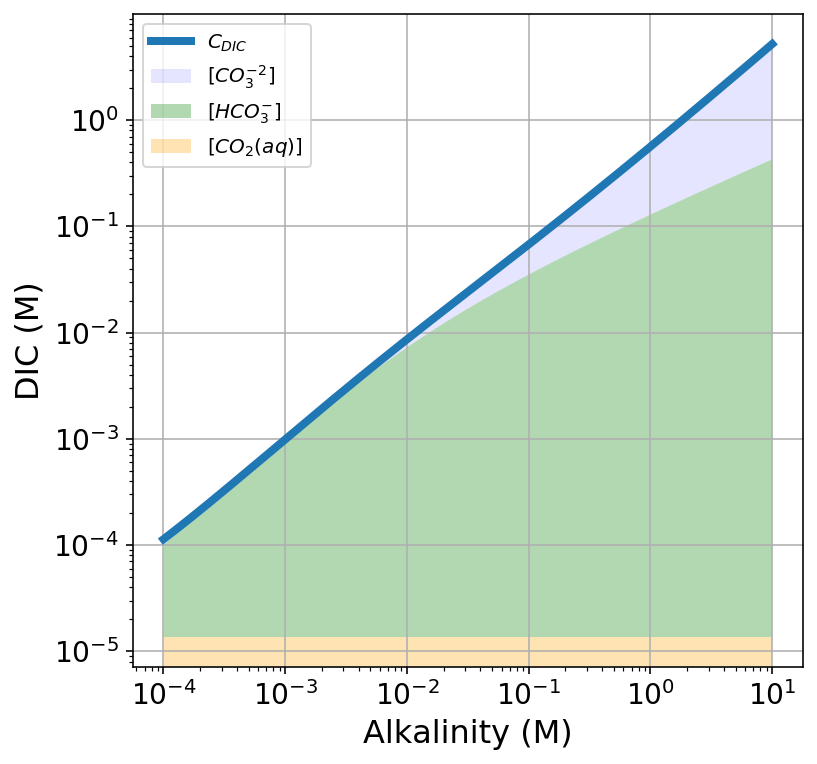}
    \caption{A plot of $C_\text{DIC}$ as a function of alkalinity at $p_\text{CO$_2$} = 0.4$ mbar; $K_1 =$ \SI{9.6e-7}{M} and $K_2 =$ \SI{3.4e-10}{M}. The heights of the three shaded areas correspond to the concentrations of aqueous carbon dioxide, bicarbonate, and carbonate molecules at a given alkalinity value.}
    \label{fig:DIC_v_Alk}
\end{figure}

Once $h$ is found, the rest of the variables are uniquely determined. Figure \ref{fig:DIC_v_Alk} shows the results of solving Equation \ref{eq:alk_pco2}, plotting $C_\text{DIC}$ as a function of alkalinity. At the values of $K_1$ and $K_2$ specified above, $b=c$ at $A\approx 10^{-2}$ M, which corresponds to $h=K_2$ or approximately pH 9.5.

\subsection{Partial pressure of \ce{CO2} as a function of alkalinity and DIC}
\begin{figure}[H]
    \centering
    \includegraphics[width=12cm]{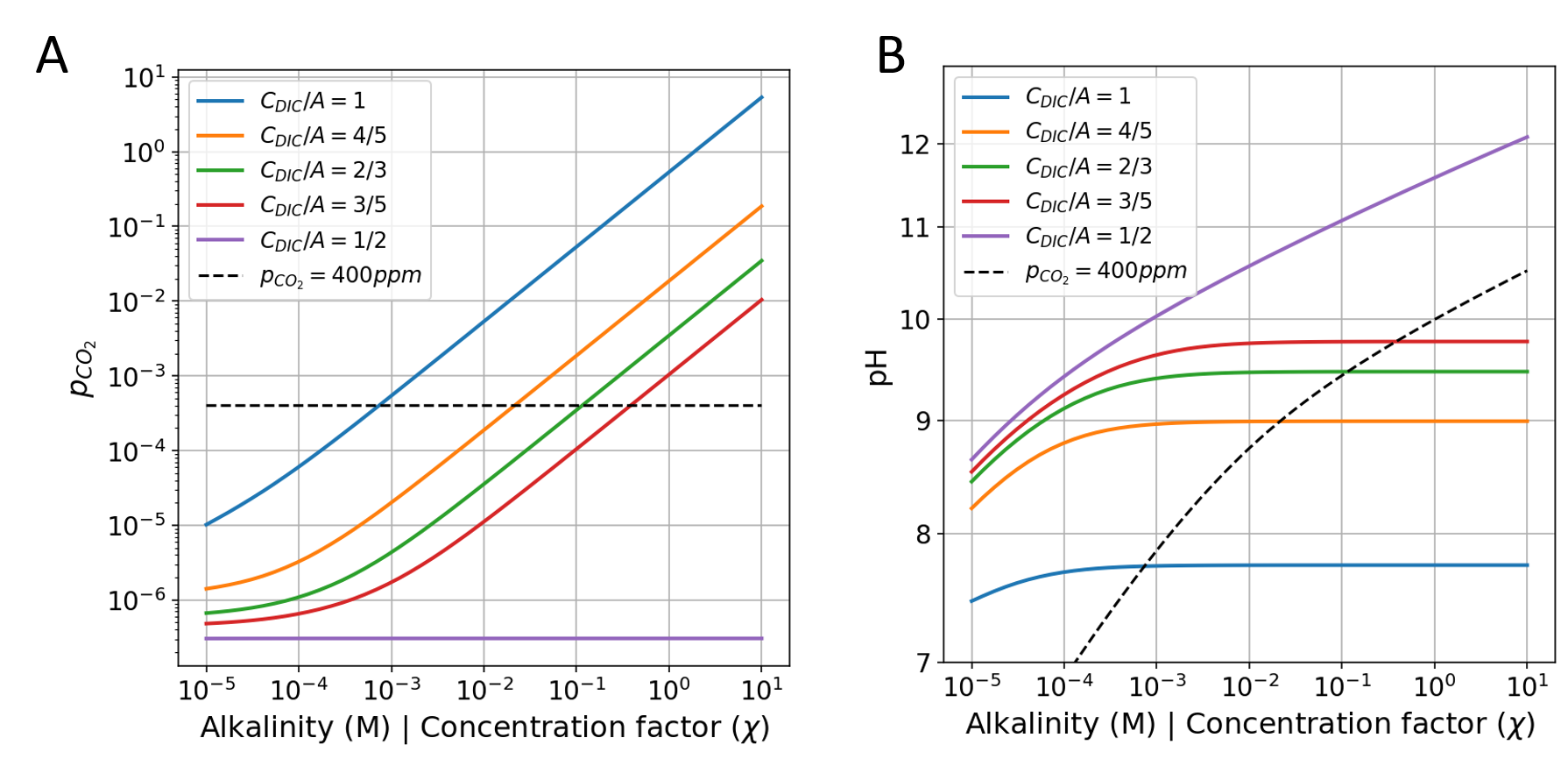}
    \caption{Concentrating alkalinity and DIC. In both panels, the x-axis plots the alkalinity or, equivalently, the concentration factor ($\chi$), given a solution initialized at 1M alkalinity. Each curve corresponds to a system with a fixed ratio between DIC and alkalinity, plotting the equilibrium $p_\text{CO$_2$}$ (A) and pH (B). Black dashed lines correspond to solution with a fixed $p_\text{CO$_2$} = 0.4$ mbar as a function of alkalinity. $C_\text{DIC}/A=1$ corresponds to a theoretical solution with only bicarbonate ions. $C_\text{DIC}/A=1/2$ corresponds to a theoretical solution with only carbonate ions.}
    \label{fig:concentrating}
\end{figure}

Here, we evaluate the case where DIC and alkalinity sets the partial pressure of \ce{CO2} ($p_\text{CO$_2$}$) of solution. The same alkalinity relationship as Eq. \ref{eq:alkalinity} applies in this condition. Given $C_\text{DIC} = a + b + c$ and the equilibrium relations from Equations \ref{eq:K1} and \ref{eq:K2}, we rewrite the alkalinity charge-balance equation as a function of $C_\text{DIC}$ and solve for $h$:
\begin{equation}
    A = C_\text{DIC} \dfrac{1 + 2 K_2/h}{1 + h/K_1 + K_2/h} + K_\text{w}/h - h
    \label{eq:fixeddic}
\end{equation}
Figure \ref{fig:concentrating} plots $p_\text{CO$_2$}$ and pH as functions of alkalinity and given a fixed $C_\text{DIC}/A$ ratio. The x-axis, equivalently, corresponds to a concentration factor ($\chi$), provided that the solution is initially equilibrated at $A=1$ M. The intersection between the curves and the dashed black line correspond to the alkalinity and pH points at which $p_\text{CO$_2$}=0.4$ mbar. This shows that a solution equilibrated at atmospheric \ce{CO2} that is concentrated with a given $C_\text{DIC}/A$ ratio increases linearly in $p_\text{CO$_2$}$ and does not change its pH. Solutions with $C_\text{DIC}/A$ ratios close to 1/2 (condition where DIC is dominated by carbonate ions) are outside of the atmospheric \ce{CO2} equilibrium conditions and do not intersect the dashed black lines or have the same partial pressure or pH scaling relations.

\section{Alkalinity concentration swing}

The alkalinity concentration swing (ACS) is based on concentrating a feed solution with a given initial alkalinity ($A_\text{i}$), which has equilibrated at a \ce{CO2} partial pressure of $p_\text{i}$, to a high alkalinity state ($A_\text{f}$). At the higher concentration, a pressure of $p_\text{f}$ is applied necessary to extract \ce{CO2} from the concentrate solution. Specifying these four parameters — $A_\text{i}$, $A_\text{f}$, $p_\text{i}$, and $p_\text{f}$ — entirely determines the outgassing properties of the ACS, based only on equilibrium assumptions. 

\subsection{ACS outgassing concentration and fraction}

The DIC concentration outgassed as \ce{CO2} ($C_\text{out}$) with respect to the feed solution is then given by:
\begin{equation}
    C_\text{out}  = C_\text{DIC}(A_\text{i},p_\text{i}) - \dfrac{A_\text{i}}{A_\text{f}} C_\text{DIC}(A_\text{f},p_\text{f})
    \label{eq:acs_out}
\end{equation}

The second term is divided by the concentration factor, $\chi = A_\text{f} / A_\text{i}$, in order to rescale the equilibrium $C_\text{DIC}$ in the concentrated state with respect to the feed solution. The fraction of DIC outgassed as \ce{CO2} with respect to input DIC is then:
\begin{equation}
    f_\text{out} = \frac{C_\text{out}}{C_\text{DIC}(A_\text{i},p_\text{i})}
\end{equation}

The equilibration between the two concentration states is driven through the following disproportionation reaction:

\begin{equation}
    \text{2HCO$_{3}^{-}$ $\rightarrow$ CO$_2$(aq) + CO$_3^{-2}$ + H$_2$O}
\end{equation}

This equation form illustrates that for each bicarbonate ion that becomes a carbon dioxide molecule, one also becomes a carbonate ion. So the value of $C_\text{out}$ is equivalent to the concentration of bicarbonates converted to carbonates. Because bicarbonate is the only species that is being converted to \ce{CO2}, Equation \ref{eq:acs_out} is equivalent to the following expression:
\begin{equation}
    C_\text{out}  = \dfrac{1}{2} \left( b(A_\text{i},p_\text{i}) - \dfrac{A_\text{i}}{A_\text{f}} b(A_\text{f},p_\text{f}) \right)
\end{equation}
The factor of 1/2 comes from the fact that half of the bicarbonates become \ce{CO2} and the other half become CO$_3^{-2}$.

\subsection{ACS $p_\text{CO$_2$}$ outgassing limit}

Taking a solution at alkalinity $A_\text{i}$ equilibrated at $p_\text{i}$ and concentrating both alkalinity and DIC concentration by a factor, $\chi$, increases partial pressure to $p_2$ (the subscript corresponds to State 2; the full cycle is described in the following subsection). 

In the regime that the ACS is expected to operate (\SI{1e-4}{M}), both [OH$^-$] and [H$^+$] are more than a factor of approximately 1900 smaller than $C_\text{DIC}$.\footnote{At $p_\text{CO$_2$}=0.4$ mbar and $A =$ \SI{1e-4}{M}, [OH$^-$] = \SI{7.6e-8}{M}, [H$^+$] = \SI{1.3e-7}{M}, while [CO$_2$(aq)] = \SI{1.36e-5}{M}, [CO$_3^{-2}$] = \SI{2.6e-7}{M}, and [HCO$_3^-$] = \SI{9.9e-5}{M}. At higher alkalinity, the $C_\text{DIC}/[\text{OH}^-]$ and $C_\text{DIC}/[\text{H}^+]$ ratios only increase.} This means that Equation \ref{eq:fixeddic} can be simplified by approximating the proton and hydroxide concentrations as 0:

\begin{equation}
    A \approx C_\text{DIC} \gamma(h)
    \label{eq:reduced_alk}
\end{equation}

Here, $\gamma(h)$ depends on $h$ and equilibrium constants. Concentrating alkalinity and DIC together scales $A \rightarrow \chi A $ and  $C_\text{DIC}\rightarrow  \chi C_\text{DIC}$. With this substitution, $\chi$ cancels out from both the left- and right-hand side of Equation \ref{eq:reduced_alk}. This means that $\gamma(h)$ is unchanged as the solution is concentrated and so $h$ (or equivalently $pH = -log_{10}(h)$) does not change either, explaining the flat pH curves (below the 400 ppm threshold curve) in the right panel of Figure \ref{fig:concentrating}. 

Using equilibrium relations, we write an expression for $p_\text{CO$_2$}$ given DIC and $h$:
\begin{equation}
    p_\text{CO$_2$} = \dfrac{1}{H^\text{cp}} \dfrac{C_\text{DIC}}{1 + K_1/h + K_1 K_2 / h^2}
\end{equation}

Because $h$ does not change as the system is concentrated ($C_\text{DIC}\rightarrow \chi C_\text{DIC}$), $p_\text{CO$_2$}$ scales linearly with the concentration factor. In the following equation, $p_\text{f,max}$ (which is the same as $p_2$) indicates the partial pressure of a solution that was initialized at $p_1$ and concentrated by a factor of $\chi$:
\begin{equation}
    p_\text{f,max} \approx \chi p_\text{i}
    \label{eq:pmax}
\end{equation}

Here, $p_\text{f,max}$ indicates the maximum \ce{CO2} partial pressure that the solution reaches after being concentrated. To extract \ce{CO2}, the solution must be exposed to a final partial pressure ($p_\text{f}$) that is lower than $p_\text{f,max}$. The left panel of Figure \ref{fig:concentrating} shows the linear scaling derived in Equation \ref{eq:pmax} for the different curves above $p_\text{CO$_2$}=400ppm$ (0.4 mbar).

\subsection{The alkalinity concentration swing cycle}
The full ACS cycle is described by four states, corresponding to four steps: concentrating (Step 1 $\rightarrow$ 2), outgassing (Step 2 $\rightarrow$ 3), diluting (Step 3 $\rightarrow$ 4), and absorbing (Step 4 $\rightarrow$ 1). Using the derived relationships and equations above, the following table details the 4 states of the ACS:
\begin{table*}[h]
\small
  \label{tbl:example}
  \begin{tabular*}{\textwidth}{@{\extracolsep{\fill}}lllll}
    \hline
    State & Alkalinity & Partial pressure & DIC &\\
    \hline
    1 & $A_1 = A_\text{i}$ & $p_1 = p_\text{i}$ (atmosphere) & $C_\text{DIC,1}$  \\
    2 & $A_2 = A_\text{f}$ & $p_2 = \chi p_1 = p_\text{f,max}$ (concentrated) & $C_\text{DIC,2}$  \\
    3 & $A_3 = A_\text{f}$ & $p_3 = p_\text{f}$ (outgassing)& $C_\text{DIC,3}$ \\
    4 & $A_4 = A_\text{i}$ & $p_4 = p_3/\chi$ (diluted)& $C_\text{DIC,4}$  \\
    \hline
  \end{tabular*}
\end{table*}

\subsection{Purity of outgassed \ce{CO2}}
\ce{CO2} outgassing purity is calculated with respect to the concentration of the other atmospheric gases dissolved in aqueous solution, \ce{N2}, \ce{O2}, and \ce{Ar}. Each gas is found at a different atmospheric partial pressure ($p_\text{N$_2$}, p_\text{O$_2$}, p_\text{Ar}$) and has a different associated Henry's constant for solubility in solution ($H_\text{i}^\text{cp}$). Given these properties, we can calculate the aqueous dissolved non-\ce{CO2} gases as:

\begin{equation}
    C_\text{non-CO$_2$} = p_\text{N$_2$}H_{\text{N}_2}^\text{cp} + p_\text{O$_2$}H_{\text{O}_2}^\text{cp} + p_\text{Ar} H_\text{Ar}^\text{cp} \approx 0.76 \text{ mM}
\end{equation}

We assume that the dissolved gases are equally dispersed through the concentrated and dilute solutions. This is because, in both RO and CDI systems we consider, dissolved aqueous gases are not affected by the respective selection mechanism, semi-permeable membrane or applied voltage. In the limiting case, all the non-\ce{CO2} gases in the concentrated stream are outgassed and collected along with the outgassed \ce{CO2} ($C_\text{out})$. We neglect the dilution of aqueous \ce{CO2} because it is found at very low concentrations ($\sim$0.014 mM). We can therefore approximate a lower bound on purity of \ce{CO2} in the following way:
\begin{equation}
    \Gamma = \dfrac{C_\text{out}}{C_\text{out} +  C_\text{non-CO$_2$}/\chi }
\end{equation}

\begin{figure}[H]
    \centering
    \includegraphics[width=12cm]{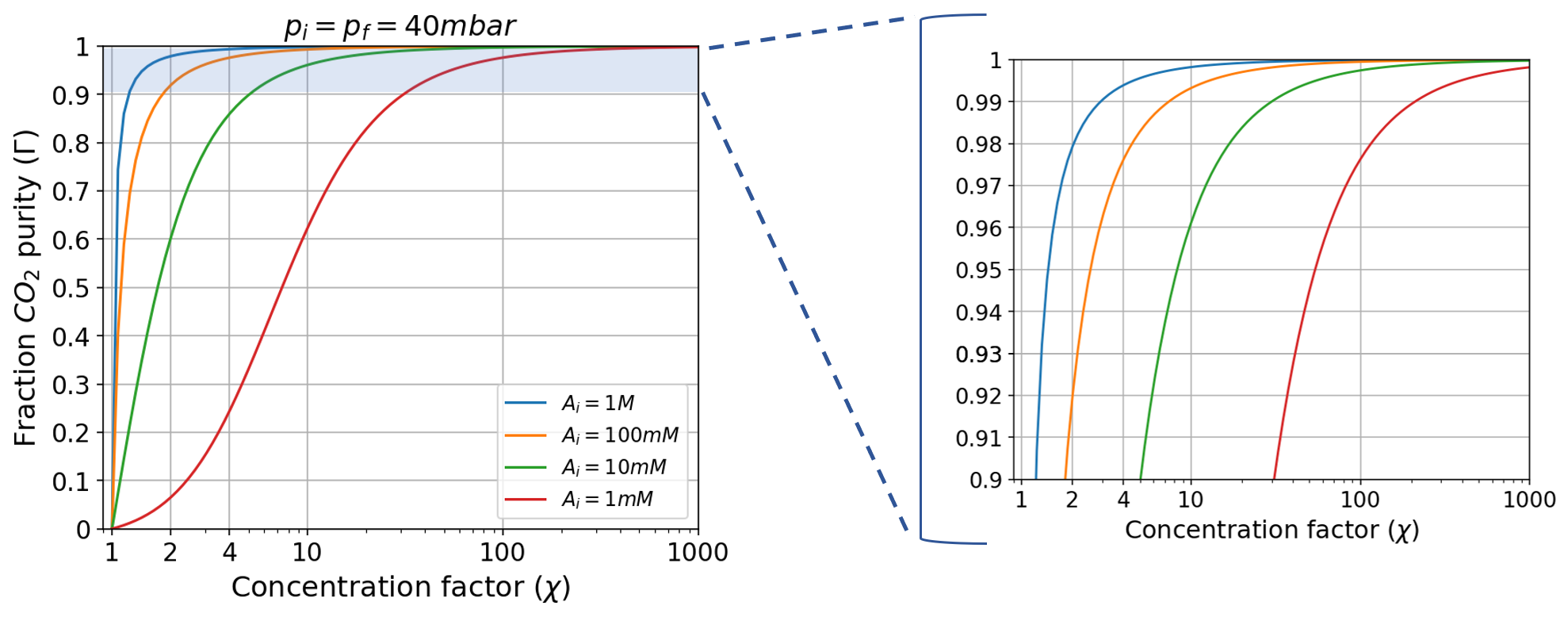}
    \caption{\ce{CO2} purity. The left panel plots the lower bound fraction of the purity of outgassed \ce{CO2} for a variety of initial alkalinity values over a range of concentration factors. The right panel is an inset of the left, showing the purity range from 0.9 to 1.}
    \label{fig:purity}
\end{figure}

Figure \ref{fig:purity} reveals that the outgassing purity depends on a relationship between the total outgassed \ce{CO2}, which depends on the initial and final alkalinity, as well as the concentration factor. In general, higher the concentration factors and higher initial alkalinity values, correspond to higher \ce{CO2} purity.

\section{Enhancement of the Concentration Swing Through Weak Bases}

We examine the case where cations in solution are a weak bases, instead of alkalinity carriers (or strong bases such as K$^+$) as in the ACS. Unlike alkalinity carriers, the dissociation state of weak bases depends on the pH and concentration. To amend the ACS thermodynamic analysis, we first write down the equilibrium properties of a weak base: 

\begin{equation}
    \text{BOH $\xleftrightarrow{K_\text{b}}$  B$^+$ + OH$^-$}
\end{equation}

This is equivalent analytically to: $BH^+ \leftrightarrow B + H^+$. If $K_\text{b}$ is the base equilibrium constant, the equilibrium relation is given by:

\begin{equation}
    K_\text{b} = \dfrac{\text{[B$^+$][OH$^-$]}}{\text{[BOH]}} = \dfrac{[\text{B}^+]K_\text{w}}{[\text{BOH}][\text{H}^+]}
\end{equation}

[B$^+$] is a function of pH and initial base concentration, $\beta_0$. Given the mass conservation identity of $\beta_0$ = [B$^+$] + [BOH], an expression for [B$^+$] can be written in the following way: 

\begin{equation}
    [\text{B}^+] = \dfrac{\beta_0}{1 + K_\text{w}/(K_\text{b} h)}
\end{equation}

We then rewrite the charge balance relation (Equation \ref{eq:alk_pco2}), as:

\begin{equation}
    [\text{B}^+] = \dfrac{\beta_0}{1 + K_\text{w}/(K_\text{b} h)} = b + 2c + K_\text{w}/h - h
\end{equation}

\begin{figure}[H]
    \centering
    \includegraphics[width=12cm]{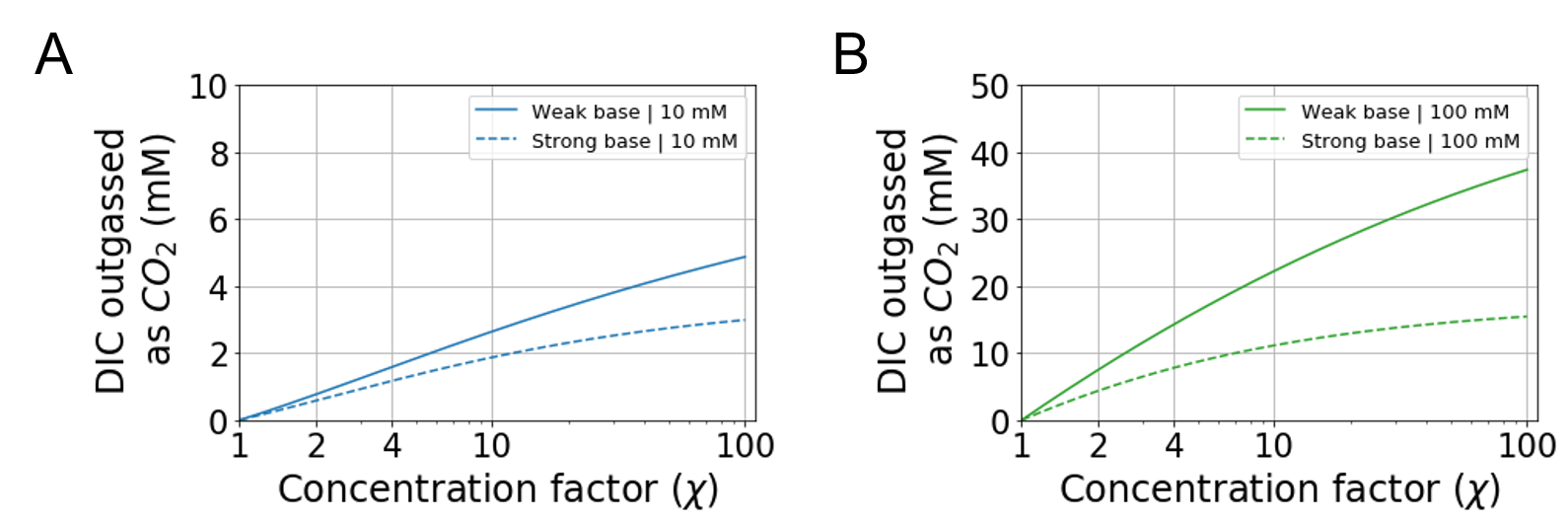}
    \caption{Concentration swing weak-base enhancement. (A) Dashed line represents a strong base (e.g. \ce{K+} or \ce{Na+}) outgassing concentration as a function of concentration factor for 10 mM initial alkalinity. Solid line represents outgassing as a result of a weak base (e.g. ammonium or ethanolamine) with $K_\text{b} = 10^{-4}$ M, initialized at 10 mM, as a function of concentration factor. (B) Same relationship as panel (A), but for initial concentration of 100 mM.}
    \label{fig:weak_base}
\end{figure}

With this modified expression, where each term is a function of $h$, we can solve for $C_\text{DIC}$ and derive modified values for $C_\text{out}$ and $p_\text{f,max}$. Figure \ref{fig:weak_base} demonstrates the enhancement in DIC outgassing caused by initializing the solution with an equivalent amount of weak base as compared to a strong base. In this example, we choose the weak base disassociation constant to be $K_\text{b} = 10^{-4}$ M, which roughly yields the largest enhancement. This value corresponds to the second disassociation constant of carbonic acid, $K_2$ (divide $K_\text{w}$ by $K_2$ to compare). The enhancement magnitude depends on the initial concentration and the concentration factor. 

\section{ACS Thermodynamic Theory and Energy Models}

\subsection{ACS thermodynamic limit}

The thermodynamic limit for any gas separation process can be set based on the ratio of the ingassing to outgassing pressures. If the ACS cycle takes an input partial pressure of \ce{CO2}, $p_\text{i}$, and outgasses at a limit output pressure of $p_\text{f,max}$, the ideal cycle work per mole \ce{CO2} is given by: $w_\text{lim} = RT \text{ln}(p_\text{f,max}/p_\text{i})$. Using Equation \ref{eq:pmax}, and rewriting in terms of the concentration factor, $\chi$, gives:

\begin{equation}
    w_\text{lim} = RT \text{ln}(\chi).
    \label{eq:thermolimit}
\end{equation}
The remainder of this section explores the work needed for vacuum outgassing, as well as two, more detailed, models that account for the energy necessary to concentrate solution and drive the ACS.
\subsection{Work needed for vacuum outgassing}

The thermodynamic minimum work needed to establish a vacuum at $p_\text{f}$ to permit \ce{CO2} outgassing isothermally is $w_\text{vac,min} = RT \text{ln}(p_\text{0}/p_\text{f})$, where $p_0 = 1$ bar. Physical systems incur additional losses from dissipation, which we model through an efficiency variable $\eta$. The real vacuum work required is then: 
\begin{equation}
    w_\text{vac} =  RT \text{ln}(p_\text{0}/p_\text{f}) / \eta
\end{equation}

\subsection{Reverse osmosis energy model}

\subsubsection{Introduction to reverse osmosis}
Reverse osmosis (RO) is a membrane-based separation method that processes an input stream of solute-filled water (feed) and creates two outputs streams, one that is more concentrated and one that is more dilute. The mechanism is driven by applying external pressure to overcome osmotic pressure generated by the difference in solute concentration across the membrane. 

RO could be applied to drive the ACS. We take a simple RO model, which assumes the Van't Hoff approximation for dilute solutions, where the pressure difference between two solutions with dissolved solutes is given by: 

\begin{equation}
    \Pi =  RT \Delta C
\end{equation}

To generalize this model, we scale this expression with $\alpha_\text{ss}$, an arbitrary dissipation term greater than 1 that corresponds to single-stage RO systems, which allows to match the RO models to physical systems:

\begin{equation}
    \Pi =  \alpha_\text{ss} RT \Delta C
\end{equation}

\begin{figure}[h]
    \centering
    \includegraphics[width=12cm]{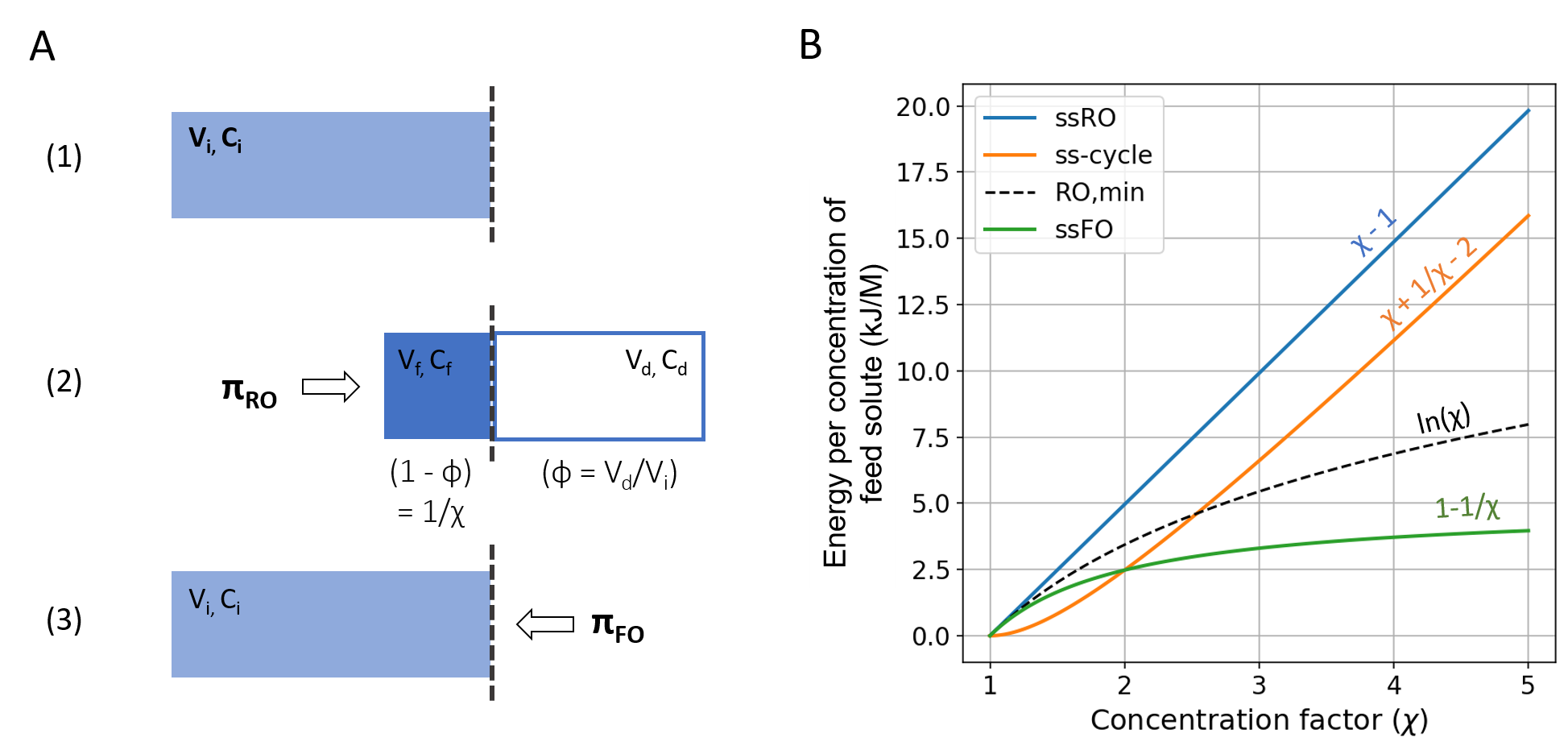}
    \caption{ Reverse osmosis model. (A) Single-stage reverse osmosis schematic. (B) A plot of energy per concentration of feed solution as a function of concentration factor for different model conditions, including single-stage RO, single-stage FO, cycle energy, and the RO thermodynamic minimum. Note that these energies are not normalized by the DIC outgassed as \ce{CO2} and are in units of energy per concentration of feed solute.}
    \label{fig:osmotic_cycle}
\end{figure}

Here, $C$ refers to the sum of the total solute concentrations, accounting for anions, cations, and non-charged molecules (e.g, [K$^+$], [CO$_2$(aq)], [HCO$_3^-]$, [CO$_3^{-2}]$). $\Pi$ is the osmotic pressure across the membrane (in units of kPa or kJ/m$^3$) and is related to the necessary applied mechanical pressure to drive the process. In this simplified model, we assume a perfectly-selective membrane that blocks all solutes and only permeates water molecules. We also assume no membrane or kinetic effects, such as concentration polarization, which will affect the necessary pressure to drive the process and corresponding energetics. We apply these assumption to an osmotic cycle (Figure \ref{fig:osmotic_cycle}A), which drives the concentrating and diluting steps of the ACS.

\subsubsection{Single-stage reverse osmosis}

First, the system is initialized with a concentration: $C_\text{i} = A_\text{i} + C_\text{DIC}(A_\text{i}, p_\text{i})$. For clarity, Figure \ref{fig:osmotic_cycle}A denotes volumes ($V_\text{i}$ = initial feed, $V_\text{d}$ = dilute solution, $V_\text{f}$ = concentrate) throughout the cycle, however, either the volume fraction ($\phi = V_\text{d} / V_\text{i}$) or concentration factor ($\chi = C_\text{f}/C_\text{i} = 1 / (1-\phi)$) of the solution is sufficient to describe the state of the system. Conceptually, in many cases, the volume fraction ($\phi$) is easier to work with for deriving the corresponding pressure and work necessary for the process.

If a concentration of $C_\text{f}$ is to be reached, the minimum applied pressure necessary is:

\begin{equation}
    \Pi_\text{RO} = \alpha_\text{ss} RT  C_\text{f}  = \alpha_\text{ss} RT  C_\text{i} / (1-\phi) = \alpha_\text{ss} RT  C_\text{i} \chi
\end{equation}

The integral over the pressure necessary to drive the system from volume fraction $0$ to $\phi$, gives the minimum work per feed volume and sets a minimum bound for the RO phase of the process. We divide the RO work by $C_\text{out}$ to get an expression for the necessary energy per mole of \ce{CO2}:
 \begin{multline}
    w_\text{RO,min} = \dfrac{\alpha_\text{ss} RT}{C_\text{out}} \int_{0}^{\phi} C_\text{i} / (1-\phi') d\phi' = -\alpha_\text{ss} RT  (C_\text{i}/C_\text{out}) \text{ln}(1 - \phi) = \alpha_\text{ss} RT (C_\text{i}/C_\text{out}) \text{ln}(\chi)
\end{multline}
 
In, what is called a “single-stage” RO process (ssRO), a single value for $\Pi_\text{RO}$ is fixed for the entirety of the concentrating phase. The single-stage RO process has the following work per volume requirement:
 \begin{equation}
    w_\text{ssRO} = \dfrac{\alpha_\text{ss} RT}{C_\text{out}} \int_{0}^{\phi} C_\text{f} d\phi' = \alpha_\text{ss} RT  C_\text{f} \phi/C_\text{out} = \alpha_\text{ss} RT (C_\text{i}/C_\text{out}) (\chi - 1)
\end{equation}

\subsubsection{Single-stage forward osmosis}

Once the input solution is concentrated and \ce{CO2} is outgassed, the system is left with two solutions, one at high solute concentration and another with pure water. It is possible to recover some of the energy stored in this concentration gradient through a forward osmosis (FO) process. An FO process recovers energy by harnessing the osmotic pressure of mixing low and high concentration solutions. Applied to the ACS, a single-stage FO process, recovers pressure at a single value, and has the following energy per volume recovery: 

\begin{multline}
    w_\text{ssFO} =\dfrac{\alpha_\text{ss} RT}{C_\text{out}} RT \int_{\phi}^{0}  (C_\text{i} - C_\text{out}) d\phi' = -\dfrac{\alpha_\text{ss} RT  (C_\text{i} - C_\text{out})}{C_\text{out}}\phi =\\
    -\dfrac{\alpha_\text{ss} RT (C_\text{i} - C_\text{out})}{C_\text{out}}(1 - 1/\chi)
\end{multline}

The adjusted solute concentration of $C_\text{i} - C_\text{out}$ comes from the fact that the solute concentration decreases after \ce{CO2} is outgassed from solution. The upper bound of energy recovery is given, as above, by: 
\begin{equation}
    w_\text{FO,max} = -\dfrac{\alpha_\text{ss} RT  (C_\text{i} - C_\text{out})}{C_\text{out}} \text{ln}(\chi)
\end{equation}

\subsubsection{Dual single-stage osmotic cycle}

Taken together, the single-stage RO concentrating and single-stage FO diluting processes, or what we call “dual single-stage” osmotic cycle, gives a total single stage cycle energy: 
 \begin{equation}
    w_\text{ss-cycle} = \alpha_\text{ss} RT  [ (C_\text{i}/C_\text{out}) (\chi - 1/\chi - 2) - (1 - 1/\chi)]
\end{equation}

Using the minimum theoretical energy for RO and maximum theoretical energy recovered for FO, we recover the same expression for energy per mole of \ce{CO2} from Equation \ref{eq:thermolimit} (given $\alpha_\text{ss} = 1$):

\begin{equation}
    (w_\text{RO,min} + w_\text{FO,max}) = RT  \text{ln}(\chi) = w_\text{lim}
\end{equation}

\subsubsection{Multi-stage reverse osmosis}

\begin{figure}[H]
    \centering
    \includegraphics[width=12cm]{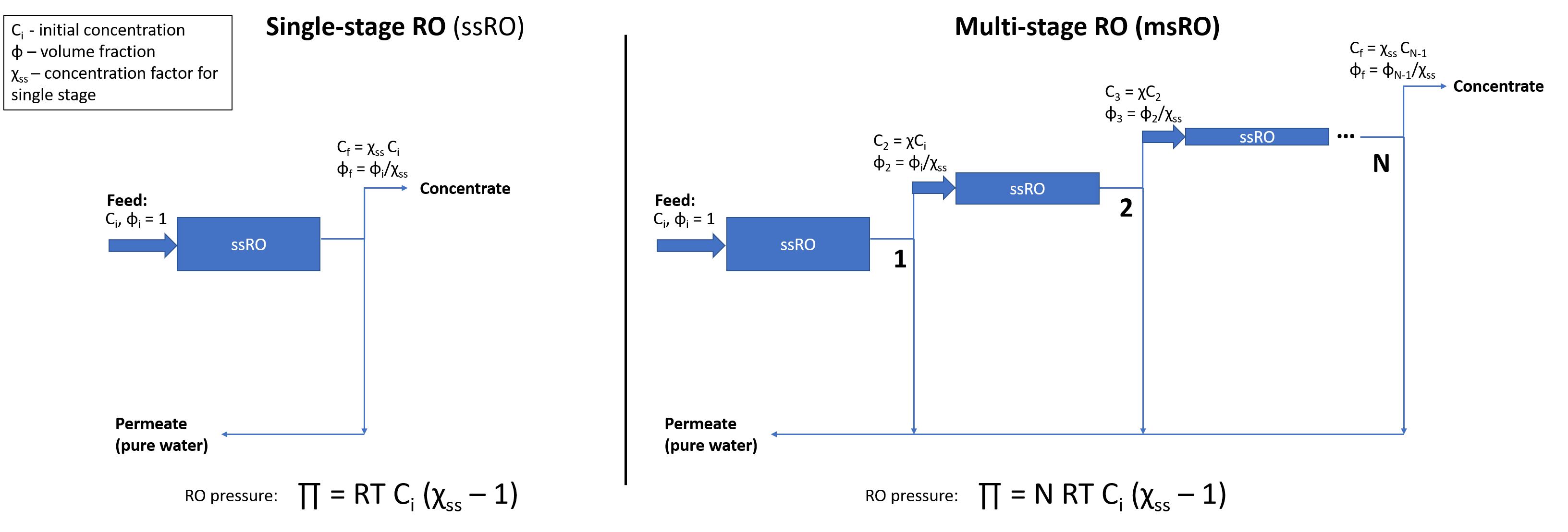}
    \caption{A schematic comparing single-stage to multi-stage RO systems.}
    \label{fig:ss_ms}
\end{figure}

In general, systems can be designed with an arbitrary number of single-stage RO modules. The more stages, the higher the energy efficiency, with the $\text{ln}(\chi)$ scaling reached at the ideal limit of infinite stages.

Figure \ref{fig:ss_ms} schematically represents the difference between single-stage and multi-stage RO (msRO) systems. Each subsequent stage along the msRO system requires processing a factor of $\chi_\text{ss}$ less feed solution, but also requires an increase in  a factor of $\chi_\text{ss}$ of applied pressure. This means that the total pressure necessary for a msRO process is:

\begin{equation}
    \Pi_\text{msRO} = \alpha_\text{ss} N RT C_\text{i} (\chi_\text{ss} -1)
\end{equation}

We again include the $\alpha_\text{ss}$ scaling term (which is $\geq$1) associated with dissipation from a single-stage module. Here N is the total number of modules in the msRO system. A total concentration factor of $\chi$ can be reached through a series of $\chi_\text{ss}$ modules: $\chi = \chi_\text{ss}^N$. We approximate the necessary number of modules through a logarithmic relation: $N \approx log_{\chi_\text{ss}}(\chi)$. (A summation series would give an exact result.) This gives the required work for \ce{CO2} extraction through a msRO process:

\begin{equation}
    w_\text{msRO} =  \alpha_\text{ss} RT (C_\text{i}/C_\text{out}) (\chi_\text{ss} -1) log_{\chi_\text{ss}}(\chi)
    \label{msRO_work}
\end{equation}

\subsection{Ion binding energy model}
We evaluate another energy model with the simple assumption that there is a single average energy cost associated with binding an ion of a given charge, independent of the concentration of ions found in the feed solution. This approximation assumes that energy scales proportionally to the alkalinity of the feed solution: $\epsilon_\text{m} A_\text{i}$. Here, $\epsilon_\text{m}$ is a characteristic energy of binding per mole, based on an ion binding energy ($\epsilon_\text{ion}$) and Avogadro's number: $\epsilon_\text{m} = \epsilon_\text{ion}N_A$. Therefore, to find a value for energy per outgassed \ce{CO2}:
\begin{equation}
    w_\text{IB} = \frac{\epsilon_\text{m} A_\text{i}}{C_\text{out}}
\end{equation}

Because the ideal limit of what the ACS can outgas is 50\% of the DIC in solution when all of the DIC is in bicarbonate form ($C_\text{out,max} = A_\text{i}/2$), the minimum of energy per mole of \ce{CO2} is: $w_\text{IB,min} = 2\epsilon_\text{IB}$.

\subsection{Energy model summary}
The following table compiles the above derivation results for the thermodynamic limit for the ACS, as well as energy per mole of \ce{CO2} separated for single-stage RO, dual single-stage cycle, multi-stage, and ion-binding models. In the RO models, the $\alpha_\text{ss}$ ($\geq1$) parameter is used to scale the model to match physical systems based on literature or experimental values.

\renewcommand{\arraystretch}{2}
\begin{center}
    \begin{tabular}{ | m{7.5em} | m{1.8cm}| m{7cm} | }
    
    \hline
    \textbf{Model} & \textbf{Variable} &  \textbf{Expression} \\  
    \hline
    Thermodynamic limit & $w_\text{lim}$ & $RT\text{ln}(\chi)$ \\ 
    \hline
    Single-stage RO  & $w_\text{ssRO}$  &  $\alpha_\text{ss} RT (C_\text{i}/C_\text{out}) (\chi - 1)$ \\ 
    \hline
    Dual single-stage cycle & $w_\text{ss-cycle}$  & $\alpha_\text{ss} RT [ (C_\text{i}/C_\text{out}) (\chi - 1/\chi - 2) - (1 - 1/\chi)]$ \\ 
    \hline
    Multi-stage RO & $w_\text{msRO}$ &  $\alpha_\text{ss} RT (C_\text{i}/C_\text{out}) log_{\chi{ss}}(\chi)(\chi_\text{ss} - 1)$ \\ 
    \hline
    Ion-binding & $w_\text{IB}$  & $\epsilon_\text{m} A_\text{i}/C_\text{out}$ \\ 
    \hline
    \end{tabular}
\end{center}
Recall that in the ACS, the concentration factor $\chi = A_\text{f}/A_\text{i}$, and the \ce{CO2} outgassing concentration ($C_\text{out}$) depends on the 4 ACS parameters: $A_\text{i},A_\text{f},p_\text{i},p_\text{f}$.
\newpage

\section{Variables}
\textbf{Carbonate chemistry:}
\begin{itemize}
  \item $C_\text{DIC}$ - Dissolved inorganic concentration (M)
  \item $[\text{CO}_2]_\text{aq} = a$ - Aqueous carbon dioxide concentration (M)
  \item $[\text{HCO}_3^-] = b$ - Bicarbonate concentration (M)
  \item $[\text{CO}_3^{-2}] = c$ - Carbonate concentration (M)
  \item $[\text{H}^+] = h$ - Proton concentration (M)
  \item $p_\text{CO$_2$}$ - Partial pressure of \ce{CO2} (bar)
  \item $K_1 =$ \SI{9.6e-7}{M} - Aqueous \ce{CO2}-bicarbonate equilibrium relation 
  \item $K_2 =$ \SI{3.4e-10}{M} - Aqueous bicarbonate-carbonate equilibrium relation
  \item $H^\text{cp} = 0.034$ M/bar - Henry's coefficient for \ce{CO2}
  \item $K_\text{w} = 10^{-14}$ \ce{M^2} - Water dissociation constant
  \item $A$ - Alkalinity (M); equivalent to [$K^+$] in this analysis unless stated otherwise
\end{itemize}
\textbf{ACS:}
\begin{itemize}
    \item $A_\text{i}$ - Initial alkalinity (M)
    \item $A_\text{f}$ - Final alkalinity (M)
    \item $p_\text{i}$ - Initial partial pressure of \ce{CO2} (bar)
    \item $p_\text{f}$ - Final partial pressure of \ce{CO2} (bar)
    \item $p_\text{f,max}$ - Maximum partial pressure of \ce{CO2} in concentrated state (bar)
    \item $C_\text{out}$ - DIC outgassed as \ce{CO2} relative to feed solution (M)
    \item $\chi = A_\text{f}/A_\text{i}$ - Concentration factor (dimensionless)
\end{itemize}
\textbf{Thermodynamics of separation:}
\begin{itemize}
    \item $R = 8.314$ J/K/mol - Gas constant
    \item $T = 293$ K - Temperature (fixed throughout entire analysis)
    \item $w_\text{lim}$ - Thermodynamic limit for gas separation process (kJ/mol)
    \item $w_\text{vac}$ - Work needed for vacuum application (kJ/mol)
    \item $\eta$ - Vacuum pump efficiency
\end{itemize}
\textbf{Reverse osmosis:}
\begin{itemize}
  \item $\Pi$ - Osmotic pressure (bar)
  \item $\Delta C$ - Difference in concentration across a semi-permeable RO membrane (M)
  \item $C_\text{i}=A_\text{i}+C_\text{DIC}(A_\text{i}, p_\text{i})$ - Initial total solute concentration (M)
  \item $C_\text{f}$ - Final total solute concentration (M)
  \item $\phi$ - Volume factor (dimensionless)
  \item $\chi = 1/(1-\phi)$ - Concentration factor (dimensionless) 
  \item $\alpha_\text{ss}$ - Single-stage RO dissipation term (greater than 1) used to rescale model to physical systems  
  \item $w_\text{ssRO}$ - Single-stage RO process work (kJ/mol\textsubscript{\ce{CO2}})
  \item $w_\text{ssFO}$ - Single-stage FO process work (kJ/mol\textsubscript{\ce{CO2}})
  \item $w_\text{ss-cycle}$ - Single-stage RO and FO cycle work (kJ/mol\textsubscript{\ce{CO2}})
  \item $w_\text{msRO}$ - Multi-stage RO process work (kJ/mol\textsubscript{\ce{CO2}})
  \item $w_\text{msRO}$ - Multi-stage RO process work (kJ/mol\textsubscript{\ce{CO2}})
\end{itemize}
\textbf{Ion binding:}
\begin{itemize}
  \item $\epsilon_\text{m}$ - Binding energy per pair of monovalent ions (kJ/mol)
  \item $w_\text{IB}$ - Energy required per mole \ce{CO2} (kJ/mol\textsubscript{\ce{CO2}})
\end{itemize}
\end{appendices}

\newpage

\bibliographystyle{unsrt}
\bibliography{references}  

\begin{thebibliography}{10}

\bibitem{minx_negative_2018}
Jan~C. Minx, William~F. Lamb, Max~W. Callaghan, Sabine Fuss, Jérôme Hilaire,
  Felix Creutzig, Thorben Amann, Tim Beringer, Wagner de~Oliveira Garcia, Jens
  Hartmann, Tarun Khanna, Dominic Lenzi, Gunnar Luderer, Gregory~F. Nemet,
  Joeri Rogelj, Pete Smith, Jose Luis~Vicente Vicente, Jennifer Wilcox, and
  Maria del Mar~Zamora Dominguez.
\newblock Negative emissions—{Part} 1: {Research} landscape and synthesis.
\newblock {\em Environmental Research Letters}, 13(6):063001, May 2018.

\bibitem{fuss_negative_2018}
Sabine Fuss, William~F. Lamb, Max~W. Callaghan, Jérôme Hilaire, Felix
  Creutzig, Thorben Amann, Tim Beringer, Wagner de~Oliveira Garcia, Jens
  Hartmann, Tarun Khanna, Gunnar Luderer, Gregory~F. Nemet, Joeri Rogelj, Pete
  Smith, José Luis~Vicente Vicente, Jennifer Wilcox, Maria del Mar~Zamora
  Dominguez, and Jan~C. Minx.
\newblock Negative emissions—{Part} 2: {Costs}, potentials and side effects.
\newblock {\em Environmental Research Letters}, 13(6):063002, May 2018.

\bibitem{ipcc_global_2018}
IPCC.
\newblock Global warming of 1.5°{C}.
\newblock Technical report, January 2018.

\bibitem{bergman_case_2021}
Andrew Bergman and Anatoly Rinberg.
\newblock The {Case} for {Carbon} {Dioxide} {Removal}: {From} {Science} to
  {Justice}.
\newblock In {\em {CDR} {Primer}}. edited by J. Wilcox, B. Kolosz, J. Freeman,
  January 2021.
\newblock https://cdrprimer.org/read/chapter-1.

\bibitem{davis_net-zero_2018}
Steven~J. Davis, Nathan~S. Lewis, Matthew Shaner, Sonia Aggarwal, Doug Arent,
  Inês~L. Azevedo, Sally~M. Benson, Thomas Bradley, Jack Brouwer, Yet-Ming
  Chiang, Christopher T.~M. Clack, Armond Cohen, Stephen Doig, Jae Edmonds,
  Paul Fennell, Christopher~B. Field, Bryan Hannegan, Bri-Mathias Hodge,
  Martin~I. Hoffert, Eric Ingersoll, Paulina Jaramillo, Klaus~S. Lackner,
  Katharine~J. Mach, Michael Mastrandrea, Joan Ogden, Per~F. Peterson,
  Daniel~L. Sanchez, Daniel Sperling, Joseph Stagner, Jessika~E. Trancik,
  Chi-Jen Yang, and Ken Caldeira.
\newblock Net-zero emissions energy systems.
\newblock {\em Science}, 360(6396), June 2018.

\bibitem{nasem_negative_2019}
NASEM.
\newblock {\em Negative {Emissions} {Technologies} and {Reliable}
  {Sequestration}: {A} {Research} {Agenda}}.
\newblock The National Academies Press, Washington, DC, 2019.

\bibitem{lenzi_ethics_2018}
Dominic Lenzi.
\newblock The ethics of negative emissions.
\newblock {\em Global Sustainability}, 1, 2018.

\bibitem{anderson_trouble_2016}
K.~Anderson and G.~Peters.
\newblock The trouble with negative emissions.
\newblock {\em Science}, 354(6309):182--183, October 2016.

\bibitem{griscom_natural_2017}
Bronson~W. Griscom, Justin Adams, Peter~W. Ellis, Richard~A. Houghton, Guy
  Lomax, Daniela~A. Miteva, William~H. Schlesinger, David Shoch, Juha~V.
  Siikamäki, Pete Smith, Peter Woodbury, Chris Zganjar, Allen Blackman, João
  Campari, Richard~T. Conant, Christopher Delgado, Patricia Elias, Trisha
  Gopalakrishna, Marisa~R. Hamsik, Mario Herrero, Joseph Kiesecker, Emily
  Landis, Lars Laestadius, Sara~M. Leavitt, Susan Minnemeyer, Stephen Polasky,
  Peter Potapov, Francis~E. Putz, Jonathan Sanderman, Marcel Silvius, Eva
  Wollenberg, and Joseph Fargione.
\newblock Natural climate solutions.
\newblock {\em Proceedings of the National Academy of Sciences},
  114(44):11645--11650, October 2017.

\bibitem{anderegg_climate-driven_2020}
William R.~L. Anderegg, Anna~T. Trugman, Grayson Badgley, Christa~M. Anderson,
  Ann Bartuska, Philippe Ciais, Danny Cullenward, Christopher~B. Field, Jeremy
  Freeman, Scott~J. Goetz, Jeffrey~A. Hicke, Deborah Huntzinger, Robert~B.
  Jackson, John Nickerson, Stephen Pacala, and James~T. Randerson.
\newblock Climate-driven risks to the climate mitigation potential of forests.
\newblock {\em Science}, 368(6497), 2020.

\bibitem{smith_soil_2016}
Pete Smith.
\newblock Soil carbon sequestration and biochar as negative emission
  technologies.
\newblock {\em Global Change Biology}, 22(3):1315--1324, 2016.

\bibitem{fajardy_can_2017}
Mathilde Fajardy and Niall~Mac Dowell.
\newblock Can {BECCS} deliver sustainable and resource efficient negative
  emissions?
\newblock {\em Energy \& Environmental Science}, 10(6):1389--1426, June 2017.

\bibitem{mcqueen_ambient_2020}
Noah McQueen, Peter Kelemen, Greg Dipple, Phil Renforth, and Jennifer Wilcox.
\newblock Ambient weathering of magnesium oxide for {CO} 2 removal from air.
\newblock {\em Nature Communications}, 11(1):3299, July 2020.

\bibitem{harvey_mitigating_2008}
L.~D.~D. Harvey.
\newblock Mitigating the atmospheric {CO2} increase and ocean acidification by
  adding limestone powder to upwelling regions.
\newblock {\em Journal of Geophysical Research: Oceans}, 113(C4), 2008.

\bibitem{rau_global_2018}
Greg~H. Rau, Heather~D. Willauer, and Zhiyong~Jason Ren.
\newblock The global potential for converting renewable electricity to
  negative-{CO} 2 -emissions hydrogen.
\newblock {\em Nature Climate Change}, 8(7):621--625, July 2018.

\bibitem{house_electrochemical_2007}
Kurt~Zenz House, Christopher~H. House, Daniel~P. Schrag, and Michael~J. Aziz.
\newblock Electrochemical {Acceleration} of {Chemical} {Weathering} as an
  {Energetically} {Feasible} {Approach} to {Mitigating} {Anthropogenic}
  {Climate} {Change}.
\newblock {\em Environmental Science \& Technology}, 41(24):8464--8470,
  December 2007.

\bibitem{renforth_assessing_2017}
Phil Renforth and Gideon Henderson.
\newblock Assessing ocean alkalinity for carbon sequestration.
\newblock {\em Reviews of Geophysics}, 55(3):636--674, 2017.

\bibitem{kelemen_overview_2019}
Peter Kelemen, Sally~M. Benson, Hélène Pilorgé, Peter Psarras, and Jennifer
  Wilcox.
\newblock An {Overview} of the {Status} and {Challenges} of {CO2} {Storage} in
  {Minerals} and {Geological} {Formations}.
\newblock {\em Frontiers in Climate}, 1, 2019.

\bibitem{sanz-perez_direct_2016}
Eloy~S. Sanz-Pérez, Christopher~R. Murdock, Stephanie~A. Didas, and
  Christopher~W. Jones.
\newblock Direct {Capture} of {CO2} from {Ambient} {Air}.
\newblock {\em Chemical Reviews}, 116(19):11840--11876, October 2016.

\bibitem{wang_moisture_2011}
Tao Wang, Klaus~S. Lackner, and Allen Wright.
\newblock Moisture {Swing} {Sorbent} for {Carbon} {Dioxide} {Capture} from
  {Ambient} {Air}.
\newblock {\em Environmental Science \& Technology}, 45(15):6670--6675, August
  2011.

\bibitem{voskian_faradaic_2019}
Sahag Voskian and T.~Alan Hatton.
\newblock Faradaic electro-swing reactive adsorption for {CO2} capture.
\newblock {\em Energy \& Environmental Science}, 12(12):3530--3547, December
  2019.

\bibitem{keith_process_2018}
David~W. Keith, Geoffrey Holmes, David St.~Angelo, and Kenton Heidel.
\newblock A {Process} for {Capturing} {CO2} from the {Atmosphere}.
\newblock {\em Joule}, 2(8):1573--1594, August 2018.

\bibitem{jin_ph_2020}
Shijian Jin, Min Wu, Roy~G. Gordon, Michael~J. Aziz, and David~G. Kwabi.
\newblock {pH} swing cycle for {CO2} capture electrochemically driven through
  proton-coupled electron transfer.
\newblock {\em Energy \& Environmental Science}, 13(10):3706--3722, October
  2020.

\bibitem{roy_dissociation_1993}
Rabindra~N Roy, Lakshimi~N Roy, Kathleen~M Vogel, C~Porter-Moore, Tara Pearson,
  Catherine~E Good, Frank~J Millero, and Douglas~M Campbell.
\newblock The dissociation constants of carbonic acid in seawater at salinities
  5 to 45 and temperatures 0 to 45°{C}.
\newblock {\em Marine Chemistry}, 44(2):249--267, December 1993.

\bibitem{ludwig_significance_2005}
Horst Ludwig and Alister~G. Macdonald.
\newblock The significance of the activity of dissolved oxygen, and other
  gases, enhanced by high hydrostatic pressure.
\newblock {\em Comparative Biochemistry and Physiology Part A: Molecular \&
  Integrative Physiology}, 140(4):387--395, April 2005.

\bibitem{zeebe_co2_2001}
Richard Zeebe and Dieter Wolf-Gladrow.
\newblock {\em {CO2} in {Seawater}: {Equilibrium}, kinetics, isotopes}.
\newblock January 2001.

\bibitem{fritzmann_state---art_2007}
C.~Fritzmann, J.~Löwenberg, T.~Wintgens, and T.~Melin.
\newblock State-of-the-art of reverse osmosis desalination.
\newblock {\em Desalination}, 216(1):1--76, October 2007.

\bibitem{elimelech_future_2011}
Menachem Elimelech and William~A. Phillip.
\newblock The {Future} of {Seawater} {Desalination}: {Energy}, {Technology},
  and the {Environment}.
\newblock {\em Science}, 333(6043):712--717, August 2011.

\bibitem{suss_water_2015}
M.~E. Suss, S.~Porada, X.~Sun, P.~M. Biesheuvel, J.~Yoon, and V.~Presser.
\newblock Water desalination via capacitive deionization: what is it and what
  can we expect from it?
\newblock {\em Energy \& Environmental Science}, 8(8):2296--2319, July 2015.

\bibitem{al-amshawee_electrodialysis_2020}
Sajjad Al-Amshawee, Mohd Yusri Bin~Mohd Yunus, Abdul Aziz~Mohd Azoddein,
  David~Geraint Hassell, Ihsan~Habib Dakhil, and Hassimi~Abu Hasan.
\newblock Electrodialysis desalination for water and wastewater: {A} review.
\newblock {\em Chemical Engineering Journal}, 380:122231, January 2020.

\bibitem{alkhudhiri_membrane_2012}
Abdullah Alkhudhiri, Naif Darwish, and Nidal Hilal.
\newblock Membrane distillation: {A} comprehensive review.
\newblock {\em Desalination}, 287:2--18, February 2012.

\bibitem{raj_review_2016}
M.~Maria~Antony Raj, K.~Kalidasa Murugavel, T.~Rajaseenivasan, and K.~Srithar.
\newblock A review on flash evaporation desalination.
\newblock {\em Desalination and Water Treatment}, 57(29):13462--13471, June
  2016.

\bibitem{shi_solar_2018}
Yusuf Shi, Chenlin Zhang, Renyuan Li, Sifei Zhuo, Yong Jin, Le~Shi, Seunghyun
  Hong, Jian Chang, Chisiang Ong, and Peng Wang.
\newblock Solar {Evaporator} with {Controlled} {Salt} {Precipitation} for
  {Zero} {Liquid} {Discharge} {Desalination}.
\newblock {\em Environmental Science \& Technology}, 52(20):11822--11830,
  October 2018.

\bibitem{boo_membrane-less_2019}
Chanhee Boo, Robert~K. Winton, Kelly~M. Conway, and Ngai~Yin Yip.
\newblock Membrane-less and {Non}-{Evaporative} {Desalination} of {Hypersaline}
  {Brines} by {Temperature} {Swing} {Solvent} {Extraction}.
\newblock {\em Environmental Science \& Technology Letters}, 6(6):359--364,
  June 2019.

\bibitem{lin_energy_2019}
Shihong Lin.
\newblock Energy {Efficiency} of {Desalination}: {Fundamental} {Insights} from
  {Intuitive} {Interpretation}.
\newblock {\em Environmental Science \& Technology}, 54(1):76--84, December
  2019.

\bibitem{jones_state_2019}
Edward Jones, Manzoor Qadir, Michelle T.~H. van Vliet, Vladimir Smakhtin, and
  Seong-mu Kang.
\newblock The state of desalination and brine production: {A} global outlook.
\newblock {\em Science of The Total Environment}, 657:1343--1356, March 2019.

\bibitem{straub_pressure-retarded_2016}
Anthony~P. Straub, Akshay Deshmukh, and Menachem Elimelech.
\newblock Pressure-retarded osmosis for power generation from salinity
  gradients: is it viable?
\newblock {\em Energy \& Environmental Science}, 9(1):31--48, January 2016.

\bibitem{willauer_recovery_2009}
Heather~D. Willauer, Dennis~R. Hardy, M.~Kathleen Lewis, Ejiogu~C. Ndubizu, and
  Frederick~W. Williams.
\newblock Recovery of {CO2} by {Phase} {Transition} from an {Aqueous}
  {Bicarbonate} {System} under {Pressure} by {Means} of {Multilayer} {Gas}
  {Permeable} {Membranes}.
\newblock {\em Energy \& Fuels}, 23(3):1770--1774, March 2009.

\bibitem{bhaumik_hollow_2004}
Debabrata Bhaumik, Sudipto Majumdar, Qiuxi Fan, and Kamalesh~K. Sirkar.
\newblock Hollow fiber membrane degassing in ultrapure water and
  microbiocontamination.
\newblock {\em Journal of Membrane Science}, 235(1):31--41, June 2004.

\bibitem{stolaroff_carbon_2008}
Joshuah~K. Stolaroff, David~W. Keith, and Gregory~V. Lowry.
\newblock Carbon {Dioxide} {Capture} from {Atmospheric} {Air} {Using} {Sodium}
  {Hydroxide} {Spray}.
\newblock {\em Environmental Science \& Technology}, 42(8):2728--2735, April
  2008.

\bibitem{wilcox_carbon_2012}
Jennifer Wilcox.
\newblock {\em Carbon {Capture}}.
\newblock Springer, January 2012.

\bibitem{sablani_concentration_2001}
SS~Sablani, MFA Goosen, R~Al-Belushi, and M~Wilf.
\newblock Concentration polarization in ultrafiltration and reverse osmosis: a
  critical review.
\newblock {\em Desalination}, 141(3):269--289, December 2001.

\bibitem{qin_comparison_2019}
Mohan Qin, Akshay Deshmukh, Razi Epsztein, Sohum~K. Patel, Oluwaseye~M.
  Owoseni, W.~Shane Walker, and Menachem Elimelech.
\newblock Comparison of energy consumption in desalination by capacitive
  deionization and reverse osmosis.
\newblock {\em Desalination}, 455:100--114, April 2019.

\bibitem{baker_membrane_2004}
Richard Baker.
\newblock {\em Membrane {Technology} and {Applications}, 2nd {Edition}}.
\newblock March 2004.

\bibitem{penate_current_2012}
Baltasar Peñate and Lourdes García-Rodríguez.
\newblock Current trends and future prospects in the design of seawater reverse
  osmosis desalination technology.
\newblock {\em Desalination}, 284:1--8, January 2012.

\bibitem{wilf_fundamentals_2014}
Mark Wilf.
\newblock Fundamentals of {RO}-{NF} technology.
\newblock {\em International conference on desalination costing, Limassol},
  pages 1--9, 2014.

\bibitem{zhao_energy_2013}
R.~Zhao, S.~Porada, P.~M. Biesheuvel, and A.~van~der Wal.
\newblock Energy consumption in membrane capacitive deionization for different
  water recoveries and flow rates, and comparison with reverse osmosis.
\newblock {\em Desalination}, 330:35--41, December 2013.

\bibitem{sarai_atab_operational_2016}
M.~Sarai~Atab, A.~J. Smallbone, and A.~P. Roskilly.
\newblock An operational and economic study of a reverse osmosis desalination
  system for potable water and land irrigation.
\newblock {\em Desalination}, 397:174--184, November 2016.

\bibitem{park_design_2020}
Kiho Park, Liam Burlace, Nirajan Dhakal, Anurag Mudgal, Neil~A. Stewart, and
  Philip~A. Davies.
\newblock Design, modelling and optimisation of a batch reverse osmosis ({RO})
  desalination system using a free piston for brackish water treatment.
\newblock {\em Desalination}, 494:114625, November 2020.

\bibitem{zhao_energy_2012}
R.~Zhao, P.~M. Biesheuvel, and A.~van~der Wal.
\newblock Energy consumption and constant current operation in membrane
  capacitive deionization.
\newblock {\em Energy \& Environmental Science}, 5(11):9520--9527, October
  2012.

\bibitem{legrand_solvent-free_2018}
L.~Legrand, O.~Schaetzle, R.~C.~F. de~Kler, and H.~V.~M. Hamelers.
\newblock Solvent-{Free} {CO2} {Capture} {Using} {Membrane} {Capacitive}
  {Deionization}.
\newblock {\em Environmental Science \& Technology}, 52(16):9478--9485, August
  2018.

\bibitem{dlugolecki_energy_2013}
Piotr Dlugolecki and Albert van~der Wal.
\newblock Energy recovery in membrane capacitive deionization.
\newblock {\em Environmental Science \& Technology}, 47(9):4904--4910, May
  2013.

\bibitem{zhang_quantifying_2020}
Yuzhong Zhang, Ritesh Gautam, Sudhanshu Pandey, Mark Omara, Joannes~D.
  Maasakkers, Pankaj Sadavarte, David Lyon, Hannah Nesser, Melissa~P.
  Sulprizio, Daniel~J. Varon, Ruixiong Zhang, Sander Houweling, Daniel
  Zavala-Araiza, Ramon~A. Alvarez, Alba Lorente, Steven~P. Hamburg, Ilse Aben,
  and Daniel~J. Jacob.
\newblock Quantifying methane emissions from the largest oil-producing basin in
  the {United} {States} from space.
\newblock {\em Science Advances}, 6(17), April 2020.

\bibitem{danckwerts_absorption_1950}
P.~V. Danckwerts.
\newblock Absorption by simultaneous diffusion and chemical reaction.
\newblock {\em Transactions of the Faraday Society}, 46(0):300--304, January
  1950.
\newblock Publisher: The Royal Society of Chemistry.

\bibitem{cong_new_2009}
Hai-Bing Cong, Ting-Lin Huang, Bei-Bei Chai, and Jian-Wei Zhao.
\newblock A new mixing–oxygenating technology for water quality improvement
  of urban water source and its implication in a reservoir.
\newblock {\em Renewable Energy}, 34(9):2054--2060, September 2009.

\bibitem{kullenberg_vertical_1976}
Gunnar E.~B. Kullenberg.
\newblock On vertical mixing and the energy transfer from the wind to the
  water.
\newblock {\em Tellus}, 28(2):159--165, April 1976.

\bibitem{fthenakis_land_2009}
Vasilis Fthenakis and Hyung~Chul Kim.
\newblock Land use and electricity generation: {A} life-cycle analysis.
\newblock {\em Renewable and Sustainable Energy Reviews}, 13(6):1465--1474,
  August 2009.

\bibitem{ahmed_hybrid_2020}
Farah~Ejaz Ahmed, Raed Hashaikeh, and Nidal Hilal.
\newblock Hybrid technologies: {The} future of energy efficient desalination
  – {A} review.
\newblock {\em Desalination}, 495:114659, December 2020.

\bibitem{tularam_environmental_2007}
Gurudeo~Anand Tularam and Mahbub Ilahee.
\newblock Environmental concerns of desalinating seawater using reverse
  osmosis.
\newblock {\em Journal of environmental monitoring: JEM}, 9(8):805--813, August
  2007.

\end{thebibliography}

\end{document}